\documentclass[apjl]{emulateapj}

\usepackage{epsfig}
\usepackage{verbatim}
\usepackage{apjfonts}

\begin{document}

\shortauthors{TREMBLAY ET AL.}
\shorttitle{ISOPHOTES \& DUST IN RADIO GALAXIES}

\slugcomment{Accepted for publication in ApJ, 17 May 2007}

\title{Isophotal  Structure   and  Dust  Distribution   in  Radio-Loud
Elliptical Galaxies}

\author{G.~R.~Tremblay\altaffilmark{1}}
\author{M.~Chiaberge\altaffilmark{1,2}}
\author{C.~J.~Donzelli\altaffilmark{1,3}}
\author{A.~C.~Quillen\altaffilmark{4}}
\author{A.~Capetti\altaffilmark{5}}
\author{W.~B.~Sparks\altaffilmark{1}}
\author{F.~D.~Macchetto\altaffilmark{1}}

\altaffiltext{1}{Space  Telescope Science  Institute, 3700  San Martin
Drive, Baltimore, MD 21218; grant@stsci.edu}

\altaffiltext{2}{On leave from INAF---Istituto di Radioastronomia, Via
P.~Gobetti 101, Bologna I-40129, Italy}

\altaffiltext{3}{IATE, Observatorio Astron\'{o}mico, UNC, Laprida 854,
C\'{o}rdoba, Argentina}

\altaffiltext{4}{Department  of Physics  and Astronomy,  University of
Rochester, 600 Wilson Boulevard, Rochester, NY 14627}

\altaffiltext{5}{INAF---Osservatorio  Astronomico  di  Torino,  Strada
Osservatorio 20, 10025 Pino Torinese, Italy}

\begin{abstract}
We investigate isophotal properties and dust morphology in the nuclear
regions of 84 radio galaxies,  imaged in the optical and near-infrared
as part of {\it Hubble Space Telescope} snapshot surveys. We present a
sample-wide  trend between  host  galaxy isophotal  structure and  the
inclination of dusty circumnuclear disks at the centers of 13 of these
objects.   We   find  that  galaxies  containing   edge-on  disks  are
invariably seen to possess  boxy isophotes, while round, face-on disks
are seen  exclusively in objects  with round or  elliptical isophotes.
Dust-rich sources  with disky isophotes  are observed only  to possess
dust in  the form  of extended filamentary  lanes, and not  in settled
distributions like disks. 
As we do not expect that edge-on and face-on disks reside in different 
populations of galaxies, we conclude that perceived isophotal boxiness 
is dependent upon the angle at which the observer views the host galaxy's 
axis of symmetry.  
We discuss our results in the context of dissipative merger scenarios, 
and infer that dusty disks primarily reside in 
old, boxy remnants of gas-poor galaxy mergers, whereas filamentary 
dust lanes reside in younger disky remnants of gas-rich mergers.  
\end{abstract}


\section{Introduction}

Many  radio-loud  elliptical  galaxies  possess $\sim  100$  pc  scale
distributions of  dust and molecular gas, often  in settled, disk-like
structures surrounding the active  galactic nucleus (AGN). While it is
widely  believed  that   dusty  circumnuclear  disks  trace  accretion
reservoirs  responsible for  feeding the  $\sim 10^9  M_{\odot}$ black
holes       at      the       centers      of       these      objects
(e.g.,~\citealt{ferrarese96,jaffe,ferrarese99}),  the  origin  of  the
dust remains  a matter of debate.   Were it native to  its host galaxy
and shed from the envelopes of  mature stars, the dust, coupled to the
stellar component of its host,  should settle into a galactic symmetry
plane over the passage of a  few hundred Myr.  There is a problem with
this scenario, however, in that  the angular momentum of the stars and
dust should  then be  nearly coincident at  all times, though  no such
correlation  has  been observed  in  active  or nonactive  ellipticals
(\citealt{goudfrooij94a},  and references therein).   This has  led to
the notion that  this dust is external in  origin, and likely acquired
through a  merger, the cooling of  hot gas, or tidal  stripping from a
dust-rich                                                     companion
(e.g.,~\citealt{goudfrooij95,vandokkum95,dekoff00,tran01}).        Dust
morphologies  other than  nuclear  disks have  also  been observed  in
early-type galaxies, ranging from amorphous clumpy patches to extended
filamentary   lanes  (e.g.~\citealt{sadler85,vandenbosch94,dekoff00}).
In the externally-acquired dust scenario, these structures are perhaps
snapshots  at  successive  stages   of  a  post-merger  dust  settling
sequence, where clumps diffuse  into lanes, and where lanes eventually
form  nuclear disks,  which are  thought to  be  end-stage equilibrium
states \citep{lauer05}.

Regardless of  origin, it is  becoming clear that these  dust features
may act as tracers of  the structure, stellar population dynamics, and
post-merger  histories of their  early-type hosts,  none of  which are
particularly well-understood.   Elliptical merger remnants,  active or
otherwise, are  far more  structurally and kinematically  complex than
originally thought, and  current understanding categorizes early-types
into  two kinematically  distinct  schemes \citep{toomre72,  searle73,
cimatti04, khochfar05}.   Elliptical galaxies classified  as ``disky''
tend to be comprised of  a rapidly rotating stellar population with an
approximately  isotropic  velocity  dispersion,  giving  rise  to  the
``pointed'' shapes  of their isophotes.  Disky ellipticals  tend to be
faint with  power-law brightness profiles, and  comprise two-thirds of
the lower end of the elliptical galaxy mass spectrum \citep{bender92}.
Conversely,   ``boxy''  ellipticals   are  characterized   by  largely
anisotropic  velocity  dispersions,  and   have  been  thought  of  as
structurally   pressure-supported  rather   than  shaped   by  stellar
rotation, as is the case with disky objects.  Boxy ellipticals tend to
be luminous  and massive, with  flat core profiles and  complex, often
triaxial stellar orbits. These properties  are thought to give rise to
the  box-like   shape  of  their   isophotes  (e.g.,~\citealt{lauer95,
faber97}).

The  dramatic  disparity between  the  properties  of  boxy and  disky
ellipticals suggests that the two classes stem from distinct formation
scenarios.   This  is  widely   supported  by  a  number  of  $N$-body
simulations of  major mergers, which  concluded that the  formation of
either  a disky  or boxy  remnant  is highly  dependent on  pre-merger
initial conditions, particularly the mass ratio, angular momentum, and
dust/gas content  of the parent  galaxies (e.g.,~\citealt{hernquist93,
limaneto95, barnes96, bekki97, khochfar05}).  The general consensus is
that, on average,  unequal-mass mergers rich in gas  and dust (``wet''
mergers) produce  disky early-type remnants,  while gas-poor (``dry'')
equal-mass mergers of high-density progenitors produce boxy galaxies.

These findings motivate the study of possible connections between dust
features   and   the   properties   of  their   underlying   starlight
distributions, as  previous works have  done.~\citet{tran01} studied a
redshift-limited  sample of 60  dust-rich early  type galaxies  (31 of
which contained  dusty disks), finding  a high tendency for  the major
axes of  dusty disks  to be aligned  with their host  galaxy isophotal
major  axes,   while  there  was  considerably  more   spread  in  the
orientation  of filamentary  lanes with  respect to  isophotes  and no
obvious trend.   The tendency for dusty  disks to align  with the host
major axis was  also observed by \citet{martel00} in  a study of seven
low-redshift radio  galaxies containing  dusty disks.  That  work also
noticed that face-on disks typically reside in ``round'' galaxies with
very little  isophotal boxiness  and ellipticity, while  edge-on disks
show a  preference for boxier  hosts.  However, the small  sample size
precluded drawing conclusions from this result.

The  dust/AGN  connection  in  radio-loud  ellipticals  is  also  well
studied.  Previous  works have found  a correlation between  dust mass
and radio  emission in these  objects (e.g.,~\citealt{deruiter02}, and
references  therein).    Furthermore,  ongoing  examinations   of  the
relationship between  radio jets and  dusty disks have noted  that the
majority of jet  axes are orthogonally oriented with  respect to dusty
disk   major   axes   (e.g.,   \citealt{capetti99,sparks00,martel00}).
Whether  or not  this apparent  trend is  real or  simply a  result of
observational   bias  (as   suggested  by   more  recent   works  e.g.
\citealt{schmitt02}),  perpendicular  jet/disk  orientations  are  not
necessarily  expected at  $\sim  100$ pc  scales.   While the  angular
momentum of the innermost accreting material should align with the jet
and the spin  of the black hole (BH), dust well  outside the BH sphere
of influence  (typically of  order $\sim 10$  pc) is dominated  by the
galactic potential.  If this apparent trend is real and not the result
of  observational   bias,  these   dusty  disks  may   trace  possible
connections  between  the structure  of  galactic equipotentials,  the
history of the merger that carved them, and the spin of the central BH
in active ellipticals.

Whatever the case, the clearly  strong motivation for studying dust in
radio-loud ellipticals has  made for a compelling field  of study that
has grown over  the past decade.  Recent {\it  Hubble Space Telescope}
({\it HST})  near-infrared imaging of  a complete sample of  3CR radio
galaxies  has granted  unprecedented views  of their  elliptical hosts
less  obscured by  extinction  from dust  \citep{madrid06,donzelli06}.
This, combined with optical {\it HST} observations of the same sample,
allows us to more closely  study connections between the properties of
dust features  and the  structure of the  radio galaxy hosts  in which
they  reside.   In  this  paper,   we  expand  upon  the  findings  of
\citet{martel00}, particularly with regards to isophotal structure and
the morphology of  nuclear dusty disks and lanes.   In \S2 we describe
our sample  and associated  observations, and in  \S3 we  describe the
reduction  of this  data  and the  extraction  of isophotal  profiles,
namely those  of the fourth Fourier  cosine (``boxiness'') coefficient
$a_4/a$  and ellipticity  $\epsilon$,  from the  iterative fitting  of
ellipses to host  galaxy isophotes.  In \S4 we  present the results of
comparing dust morphology, distribution,  and inclination (in the case
of  dusty disks)  to isophotal  boxiness and  ellipticity, as  well as
radio fluxes from  the core and extended regions of  the host.  In \S5
we  discuss the  implications of  our results  in the  context  of the
expression of isophotal boxiness and diskiness in early-type galaxies.
We also  discuss the  possible origins of  dusty disks and  lanes with
regards to merger scenarios.  We summarize this work and discuss areas
of future  study in \S6.  Throughout this  paper we use $H_0  = 71$ km
s$^{-1}$ Mpc$^{-1}$, $\Omega_M = 0.27$, and $\Omega_{\Lambda} = 0.73$.

\begin{figure*}
\plotone{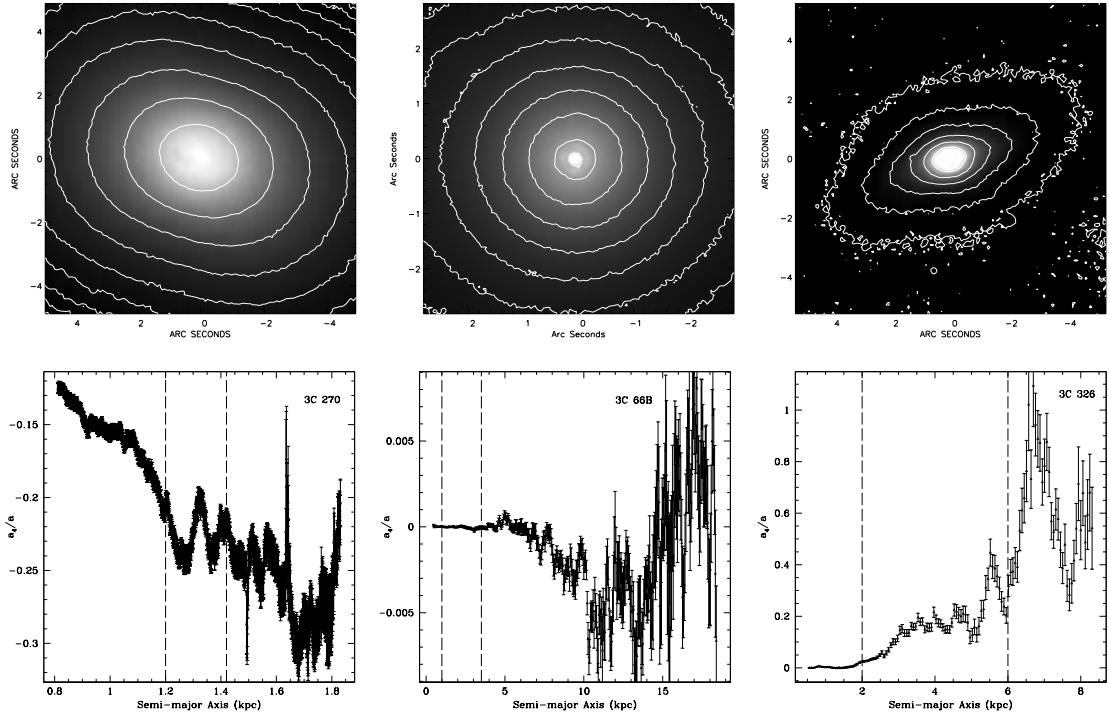}
\caption{Examples  of  isophotal boxiness  in  varying degrees.   {\it
Left}: {\it HST}/NIC2  image of 3C~270 at 1.6  $\mu$m with highlighted
isophotal contours  ({\it top}), its associated  $a_4/a$ profile ({\it
bottom}). Dashed vertical lines indicate the region over which $a_4/a$
and $\epsilon$  were averaged  for 3C~270.   The isophotes  of this
object are classified as extremely  boxy.  {\it Middle}: Same as ({\it
Left}), but for  3C~66B.  The isophotes of this  object are classified
as round.   {\it Right}:  Same as ({\it  Left}), but for  3C~326.  The
isophotes of this object are classified as extremely disky.  }
\label{fig:examples}
\end{figure*}

\section{Sample Selection and Observations}

Our study  is largely  based on  a complete redshift  ($z <  0.3$) and
flux-limited sample of Fanaroff \&  Riley (1974, hereafter FR) class I
and  II radio galaxies,  the majority  of which  are derived  from the
extragalactic  subset of  the  Revised Third  Cambridge Catalog  (3CR;
\citealt{bennet62a,   bennet62b,   spinrad85}).    3CR   sources   are
categorized as such given a flux density at 178 MHz exceeding 9 Jy, as
well as  a location on the  sky constrained to  declinations $\delta >
-5^\circ$  and galactic  latitudes  $b >  10^\circ$.  Observations  of
radio galaxies at 178 MHz  detect extended, unbeamed jet lobes, so the
sample  is  free  from  orientation  bias, making  it  well-suited  to
statistical studies of host galaxy properties.

Of the 116 3CR sources with $z  < 0.3$, our sample is comprised of the
83  for  which  there  is  archival near-infrared  imaging  from  {\it
HST}. We also include NGC 6251 in our sample, not part of the original
3CR    catalog   because    of   a    minor    position   technicality
\citep{waggett77}.  \citealt{laing83} later  included NGC 6251 as part
of their complete  revised sample (the 3CRR catalog).  This brings our
sample to  a total of 84 targets.   As this study aims  to compare the
morphology  of dust features  with the  isophotal properties  of their
hosts,  we focus  primarily on  a 33  object dust-rich  subset  of our
sample,  identified as  such  by previous  {\it  HST} optical  studies
(e.g., \citealt{dekoff96,martel99,dekoff00}). The dust features of NGC
6251  were studied  in detail  by \citet{ferrarese99},  and references
therein.   We  include the  52  remaining  dust-poor  targets for  the
purposes  of comparing  the  global isophotal  properties  of a  large
number of objects.

We obtain the archival near-infrared (NIR) and optical imaging for our
sample, allowing us to capitalize on the varying effects of extinction
from dust at  these two wavelength regimes.  As  extinction is reduced
at near-infrared compared to optical wavelengths, our analyses of host
galaxy  isophotal  properties are  largely  based  on a  near-infrared
snapshot survey  of 3CR targets at  1.6 $\mu$m (similar  to $H$ band).
These  observations were  performed  by the  Near-Infrared Camera  and
Multiobject Spectrograph (NIC2) on board {\it HST}, under SNAP program
10173  (PI:  Sparks, \citealt{madrid06}).   The  completeness of  this
program was subject  only to the observing schedule  of {\it HST}, and
not  for any  other reason  specific to  any individual  target.  This
subsample of NICMOS  NIR imaging is therefore as  bias-free as the 3CR
sample itself.  For  a small number of the objects  in our sample, the
NICMOS imaging  we utilize  was taken as  part of other  programs, the
references for which may be found in Table~\ref{tab:tab1}.

These data were  obtained from the Multimission Archive  (MAST) at the
Space Telescope  Science Institute (STScI), having  passed through the
{\it  HST}/NICMOS on-the-fly  re-processing  (OTFR) pipeline  yielding
nearly science-ready  data.  Post-pipeline corrections  (i.e., removal
of  the NICMOS  pedestal effect,  \citealt{noll04}) were  performed by
\citet{madrid06}  and  \citet{donzelli06},   in  which  more  detailed
information specific to  the NICMOS data reduction can  be found.  All
NICMOS  observations presented  in this  paper had  a  projected pixel
scale of $0\arcsec.075$, and were performed using the F160W (analog of
$H$ band)  filter over  a total exposure  time of either  1152 seconds
(for all program 10173 targets) or for slightly shorter times (see the
references in  Table~\ref{tab:tab1} for information  specific to these
images).

For nearby ($z  < 0.1$) galaxies, the {\it HST}/NIC2  field of view is
prohibitively  small and  limits our  ability to  sample near-infrared
background levels.  We therefore  utilize a companion ground-based NIR
dataset  \citep{marchesini05},  obtained   with  the  {\it  Telescopio
Nazionale Galileo} ({\it TNG}).   The observations were performed with
two  {\it TNG}  instruments over  two observing  runs. In  both cases,
imaging  was  performed with  filters  centered  around $2.12$  $\mu$m
($K'$-band) for  a typical  exposure time of  $\sim 10 -  20$ minutes,
depending  on  the  magnitude  of the  source.   Detailed  information
regarding    the   observations    is    found   in    Table   3    of
\citet{marchesini05}.

As  this  paper traces  a  connection  between  host galaxy  isophotal
properties  (studied in  the near-infrared)  and dust  content (better
observed in the optical), we retrieve archival {\it HST}/WFPC2 imaging
for those  3CR radio galaxies already  known to contain  dust lanes or
dusty disks,  as detailed by \citet{dekoff00}.  This  optical data was
retrieved  from  MAST  and  sent  through  the  {\it  HST}/WFPC2  OTFR
pipeline.   As our  use  for the  optical  data only  extended to  the
creation of color maps (discussed  in \S 3.2), we were only interested
in  the dusty  central  regions of  the  host galaxy.   As such,  OTFR
calibrated images are adequate  for our purposes, and no post-pipeline
corrections were  necessary.  Point-spread function  (PSF) differences
between  the $R$-  and $H$-band  data were  ignored, as  any variation
would only  strongly affect regions of high  surface brightness (e.g.,
very near the  nucleus, which we are not interested  in).  We were not
concerned with optical  or NIR nuclei corrupting the  natural shape of
isophotes or  color map  isochromes as they  are faint  throughout our
sample.  All {\it HST}/WFPC2 images presented in this paper were taken
with  the  F702W filter  (analog  of $R$  band),  for  which the  mean
wavelength is $\left<  \lambda \right> = 6818$ \AA.   While one of the
most  efficient configurations  for  WFPC2, it  also  includes in  its
passband  contributions  from line  and  continuum  emission.  At  the
redshifts considered  in this  paper, these include  H$\alpha$, [N{\sc
ii}], and [O{\sc iii}]. The WFPC2 Planetary Camera has a field of view
of  $36\arcsec.4 \times 36\arcsec.4$  and a  projected pixel  scale of
$0\arcsec.0455$.

\section{Radial Brightness Profiles: Quantifying Boxy and Disky Isophotes}

\begin{figure*}
\plottwo{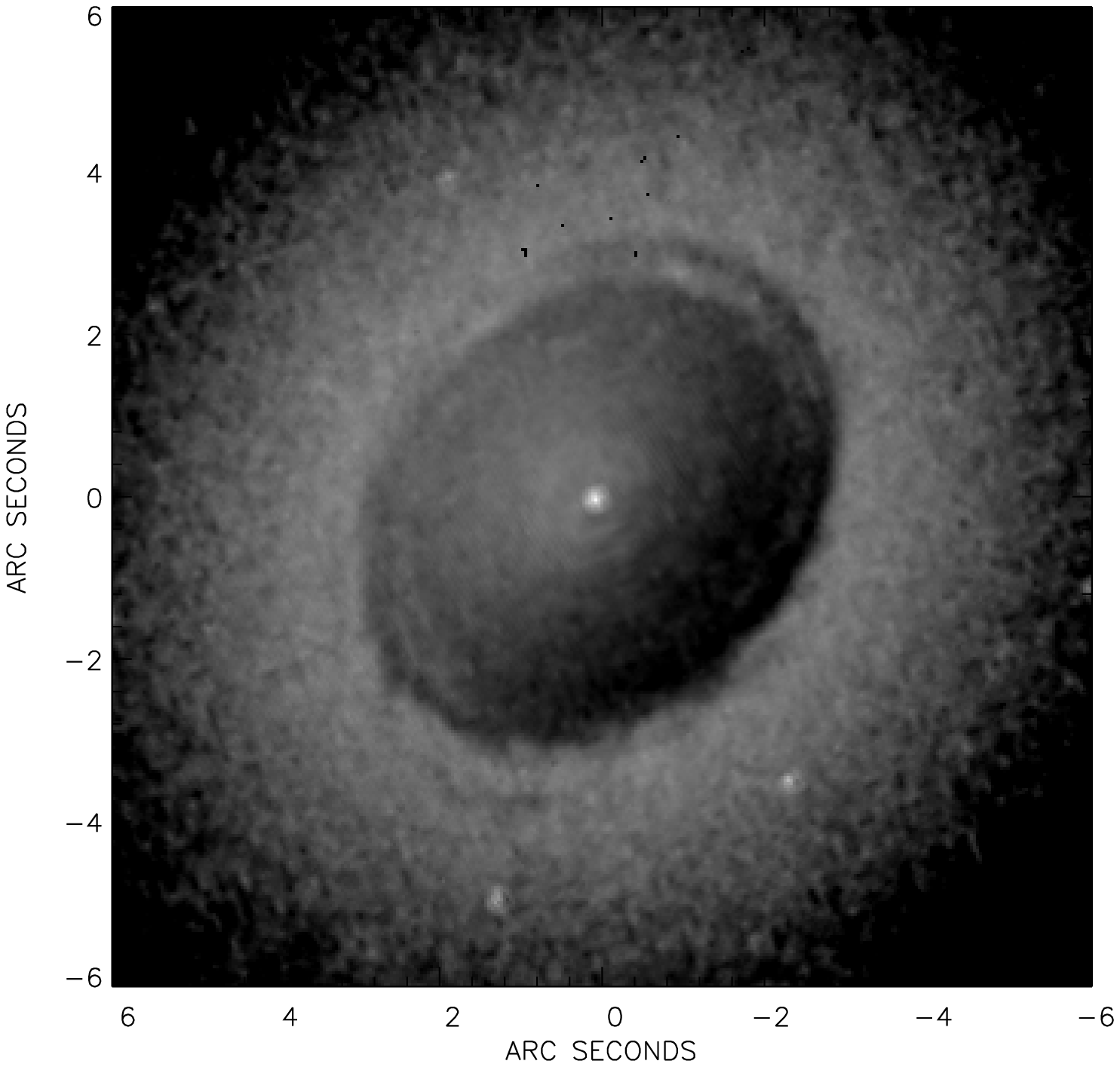}{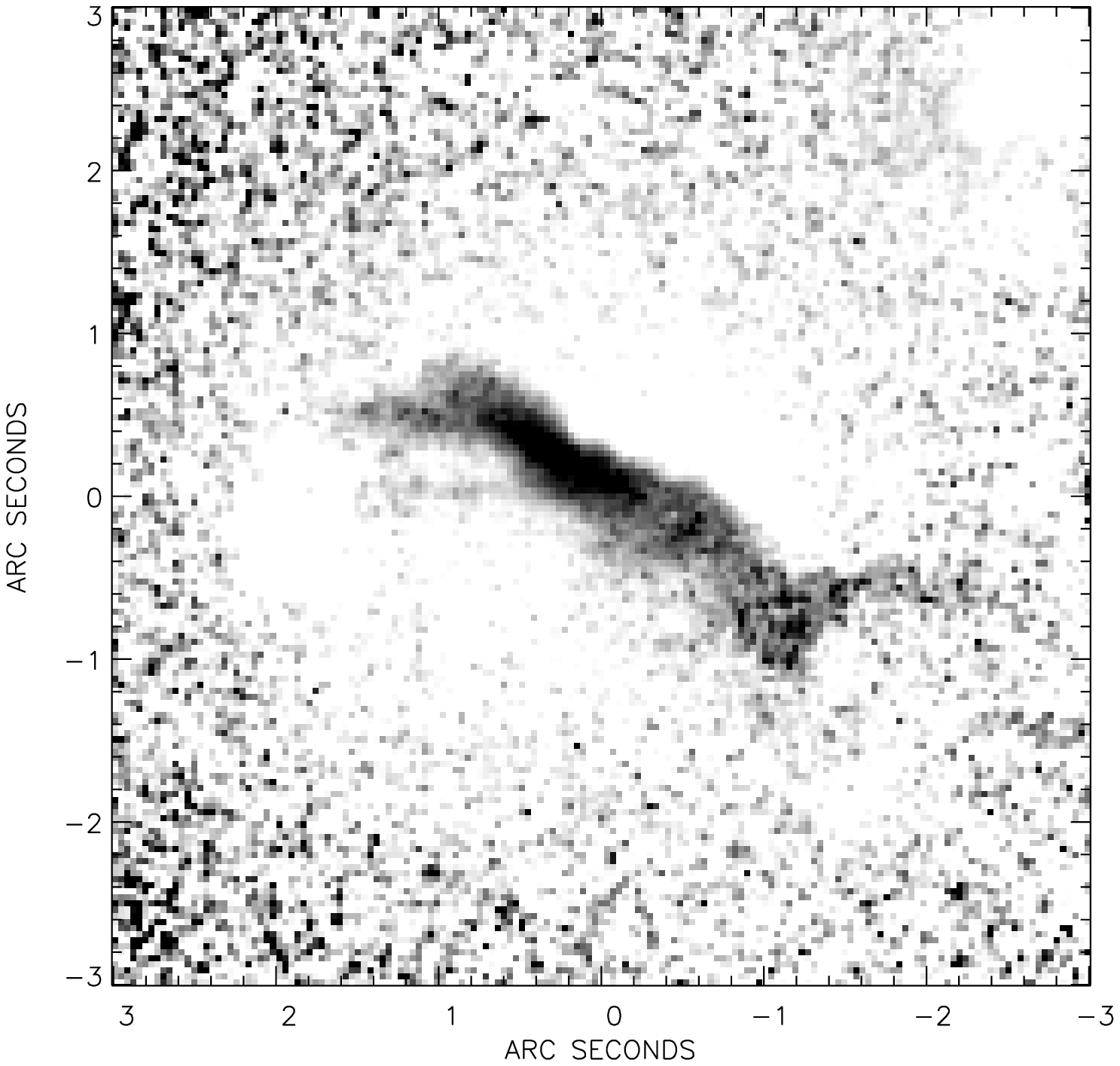}
\caption{Examples  of a  circumnuclear  dusty disk  and  a dust  lane,
respectively.   ({\it a})$1.6$ $\mu$m/0.702  $\mu$m absorption  map of
the nuclear  disk of  3C~31, made from  dividing {\it  HST}/NICMOS and
{\it  HST}/WFPC2 images.   The disk  exhibits more  absorption  on its
southwestern edge, and previous studies of the axial ratio $\left( b/a
\right)$  for the  disk have  estimated  the southwestern  edge to  be
nearer the  observer than the northeastern edge,  with $\arccos \left(
b/a \right)  \approx 41^\circ$  with respect to  the plane of  the sky
\citep{fraixburnet91,  dekoff00} ({\it  b}) $1.6$  $\mu$m/0.702 $\mu$m
absorption map of the $\sim 6.5$  kpc dust lane in 3C~321.  Note that,
while we  include 3C~321  in this figure  because it  best illustrates
what we define  as a dust lane, we  do {\it not} include it  in our 84
object sample as its $H$-band isophotes are very highly distorted by a
foreground object.  }
\label{fig:disklane}
\end{figure*}

\begin{figure}
\plotone{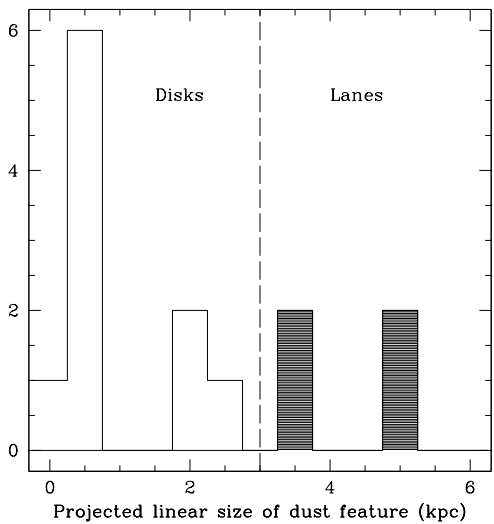}
\caption{Distribution of projected linear  sizes of the 17 dusty disks
and lanes in our sample.   The grey region of the histogram represents
dust lanes,  the white  region indicates that  the feature is  a disk.
See  Table~\ref{tab:tab1}  for a  qualitative  classification of  dust
morphology (where applicable) for every object in our sample.  }
\label{fig:radhist}
\end{figure}

Luminosity profile extraction from  the {\it HST}/NICMOS and {\it TNG}
images    via   isophotal   ellipse    fitting   was    performed   by
\citet{donzelli06} using the IRAF  routine {\sc ellipse} within STSDAS
\citep{jedrzejewski87}.  These fits were performed from the very inner
regions of  the {\it HST}  and {\it TNG}  images to the  outer regions
with  a count  level  of $  2  \sigma_{\mathrm{sky}}$.  More  detailed
information  regarding  these  fits,  including the  masking  of  dust
features   and  foreground   objects,   can  be   found   in  \S4   of
\citet{donzelli06}.

The primary  output of {\sc ellipse}  is a table  detailing the radial
profiles  of a number  of isophotal  parameters (and  their associated
uncertainties),  one of which  is the  harmonic amplitude  $B_4$.  The
standard metric for ``boxiness'' and ``diskiness'' in isophotes is the
4$^{\mathrm{th}}$   order  Fourier   cosine  coefficient   $a_4/a$,  a
measurement of  isophotal radial  deviation ($a_4$) normalized  to the
semi  major-axis $a$  at  which the  ellipse  was fit.   $B_4$ can  be
converted   to   $a_4/a$  following   the   convention  described   by
\citet{milvang-jensen99}.   That  is, {\sc  ellipse}  table values  of
$B_4$ are  normalized to  both the equivalent  radius $r  = \sqrt{ab}$
(where $a$ and  $b$ are the isophotal semi-major  and semi-minor axes,
respectively) as  well as  the local gradient  in intensity  $I$ (also
computed from {\sc ellipse}),
\begin{equation}
a_4/a  =  \frac{B_4}{r  \cdot  \left|  \frac{\mathrm{d}I}{\mathrm{d}r}
\right|} \cdot \sqrt{\frac{b}{a}}.
\end{equation}
The  extra  factor  of  $\sqrt{b/a}  =  \sqrt{1  -  \epsilon}$  (where
$\epsilon$ is the isophotal  ellipticity from {\sc ellipse}) is needed
to re-normalize the  radial deviation to $a$ as opposed  to $r$, so as
to   be    consistent   with   the   standard    in   the   literature
\citep{bender87,milvang-jensen99}.    Negative   values   of   $a_4/a$
indicate that  the host's isophotes  are boxy, while  largely positive
values of  $a_4/a$ are manifestations of  disky isophotes.  Typically,
$\left| a_4/a  \right|$ does  not exceed $\sim  0.3$.  An  object with
values  of $a_4/a \approx  -0.2$ is considered  extremely boxy
(see  Fig.~\ref{fig:examples}),  and  a  galaxy whose  isophotes  have
values around  $a_4/a \approx 0.2$  is extremely disky.   Objects with
$a_4/a$ values very close to 0 would be described as ``elliptical'' or
``round'',  as would  be the  case  for an  object with  a very  small
$a_4/a$ as well as negligible ellipticities $\epsilon$.

We  define the  average  $a_4/a$ of  a  selected range  of a  galaxy's
luminosity  profile as  $\left(a_4/a\right)_{\mathrm{ave}}$.   As this
paper  places great importance  on $\left(a_4/a\right)_{\mathrm{ave}}$
as a  physical quantity, we  must be careful  in how we  compute this
average.  For  each object, we  plot the radial dependence  of $a_4/a$
and associated errors so as  to judge its behavior at varying galactic
radii.   Based on  these plots,  we make  a selection  for  the radial
region over which we calculate $\left(a_4/a\right)_{\mathrm{ave}}$ and
$\epsilon_{\mathrm{ave}}$  based on  the  following {\it  qualitative}
constraints:
\begin{itemize}
\item The region is sufficiently large (e.g. at least $\sim 1$ kpc) to
ensure     that     the    $\left(a_4/a\right)_{\mathrm{ave}}$     and
$\epsilon_{\mathrm{ave}}$  quantities are generally  representative of
the host galaxy's overall isophotal structure.
\item In the  case of objects with nuclear  dust features (e.g. disks)
or  an unresolved  nuclear peak,  we are  careful to  avoid  the inner
regions of the host, and  begin our selection region at least 2\arcsec
away from  the nucleus. This was  done for nearly every  object in our
sample.
\item  The $a_4/a$  values  within  the selected  region  do not  have
associated uncertainties  greater than  the overall errors  typical to
the  fit.   While  this  is  a qualitative  assessment  of  error,  we
strengthen the confidence  in our selection by also  ensuring that the
mean  position  angle (PA)  of  each  ellipse  does not  vary  greatly
throughout  the region,  and that  the $a_4/a$  profile  is relatively
stable (i.e.  it does  not ``spike'' dramatically within the selection
region)
\end{itemize}

\begin{figure*}
\plottwo{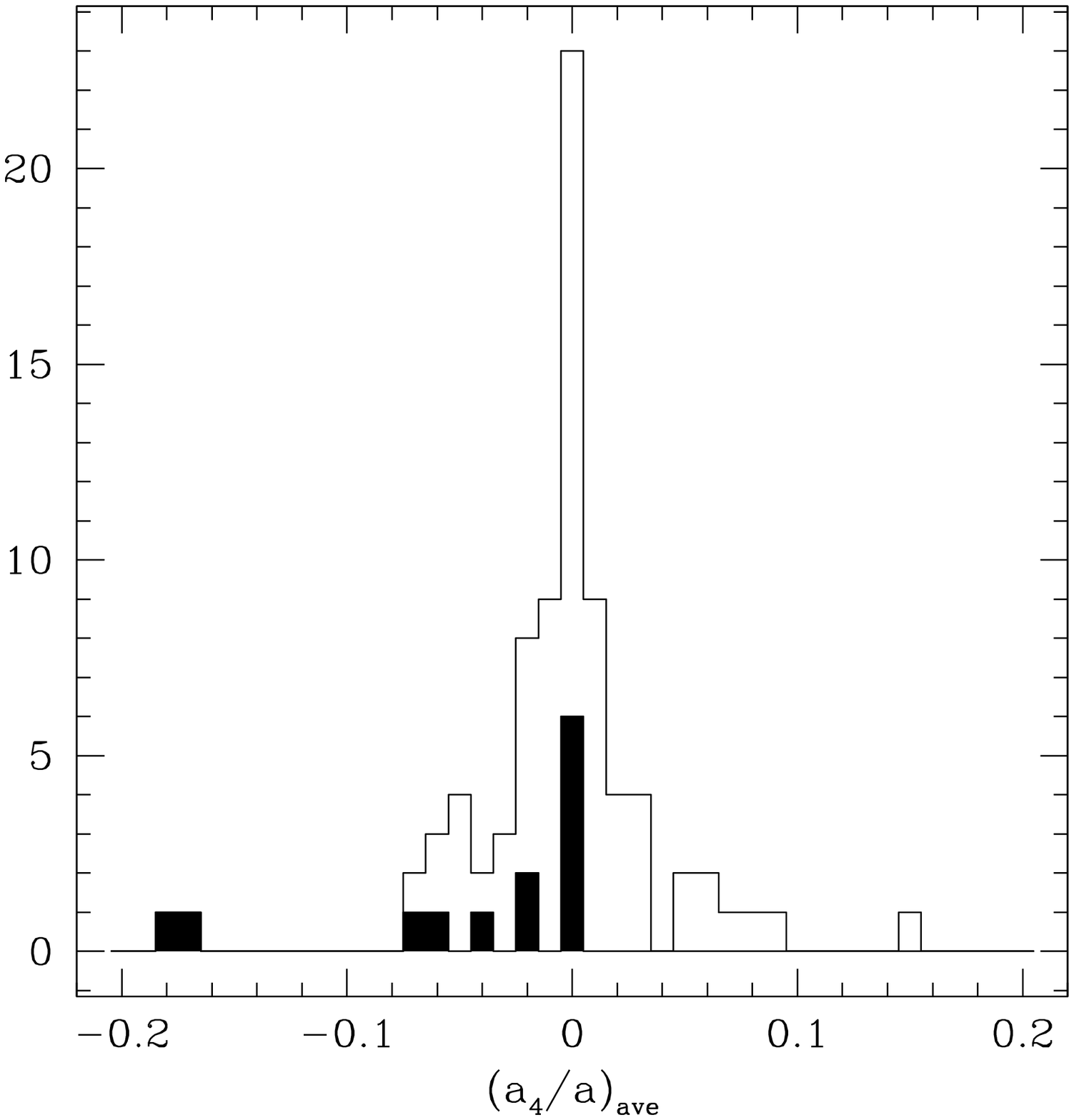}{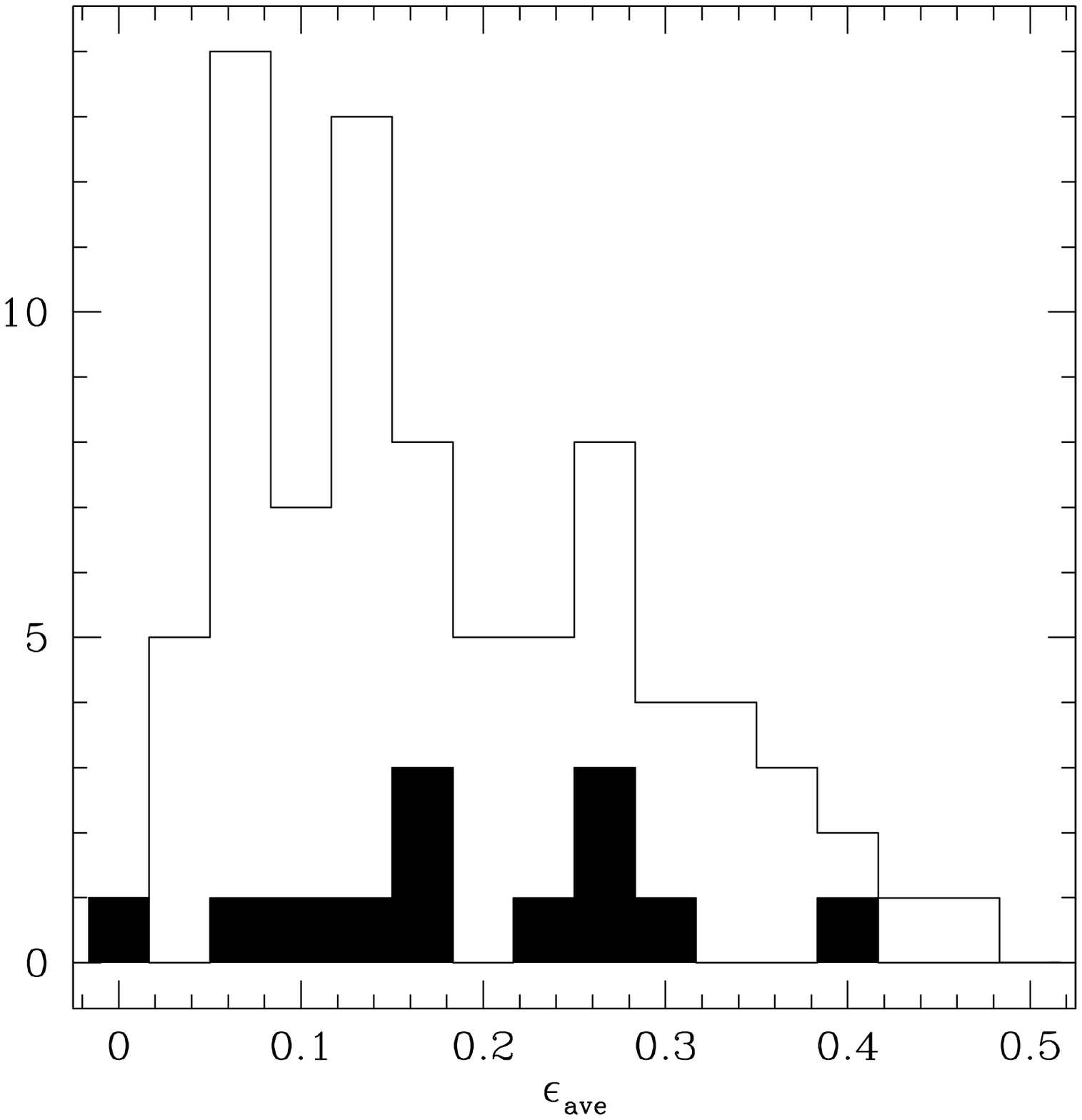}
\caption{({\it         a})         Distribution         of         the
$\left(a_4/a\right)_{\mathrm{ave}}$   boxiness   parameters  for   our
sample.   Shaded  parts  of   the  histogram  represent  the  boxiness
distribution   for  objects   containing  dusty   disks.    ({\it  b})
Distribution of mean isophotal  ellipticities for our sample, averaged
over the same  range used to average $a_4/a$ values. See  \S 3.2 for a
discussion of our strategy in selecting these ranges.  }
\label{fig:isohist}
\end{figure*}

These regions  must be selected  uniquely for each target,  as $a_4/a$
profiles  vary greatly  throughout  our sample  given  the effects  of
redshift,   obscuring  dust   features,   foreground  objects,   close
companions,  {\sc  ellipse}  fit  parameters, etc.   In  general,  the
regions over which the  above requirements were satsified fell between
the isophote $\sim 1$ kpc  from the nucleus and the galaxy's effective
radius (e.g.,  the radius  of the isophote  containing half  the total
luminosity  of the  host, see  Table 1).   This averaging  strategy is
illustrated in  Fig.~\ref{fig:examples}, in which we  give examples of
(from left to right in  the figure) very boxy (3C~270), round (3C~66B)
and very  disky (3C~326)  host galaxies, respectively.   Alongside the
{\it  HST}/NIC2 1.6  $\mu$m isophotal  contour plots  (top)  for these
objects, we  present the  associated $a_4/a$ radial  profiles (bottom)
from the {\sc  ellipse} fits.  The vertical dashed  lines on the plots
mark      the      region       over      which      we      calculate
$\left(a_4/a\right)_{\mathrm{ave}}$    and   $\epsilon_{\mathrm{ave}}$
values  for  each  host.    Eleven  of  our  luminosity  profiles  had
associated errors that  were higher than the global  average error for
our   sample.   Typically,   this  was   due  to   some  observational
irregularity, often  resulting in the  object being close to  the NIC2
chip  edge. In  other cases,  large amounts  of extinction  from dusty
clumps  and  patches  impeded  our  ability  to  confidently  fit  the
isophotes.  While  we provide  estimated isophotal properties  for the
majority of  these objects,  we do not  include them in  our following
global comparisons  of isophotal properties.   Information specific to
the  isophotes  of  every  object  in  our  sample  may  be  found  in
Table~\ref{tab:tab1}.

\section{Results}

\subsection{Dusty disks and lanes}

Of the 84 galaxies in our  sample, we identify 33 with dust in various
distributions, including  clumpy patches not centered  on the nucleus,
lanes  at  kpc scales,  and  circumnuclear  disks  hundreds of  pc  in
diameter. We  find 13 such  dusty disks, clearly resolvable  either in
optical WFPC2  data or with an  $H/R$-band color map.   Our color maps
have been normalized  such that a measured difference  in pixel values
between  two regions  corresponds to  a difference  in  a proportional
number  of $R$-band magnitudes. This  allows for  quick extinction
estimates.

\begin{figure*}
\plottwo{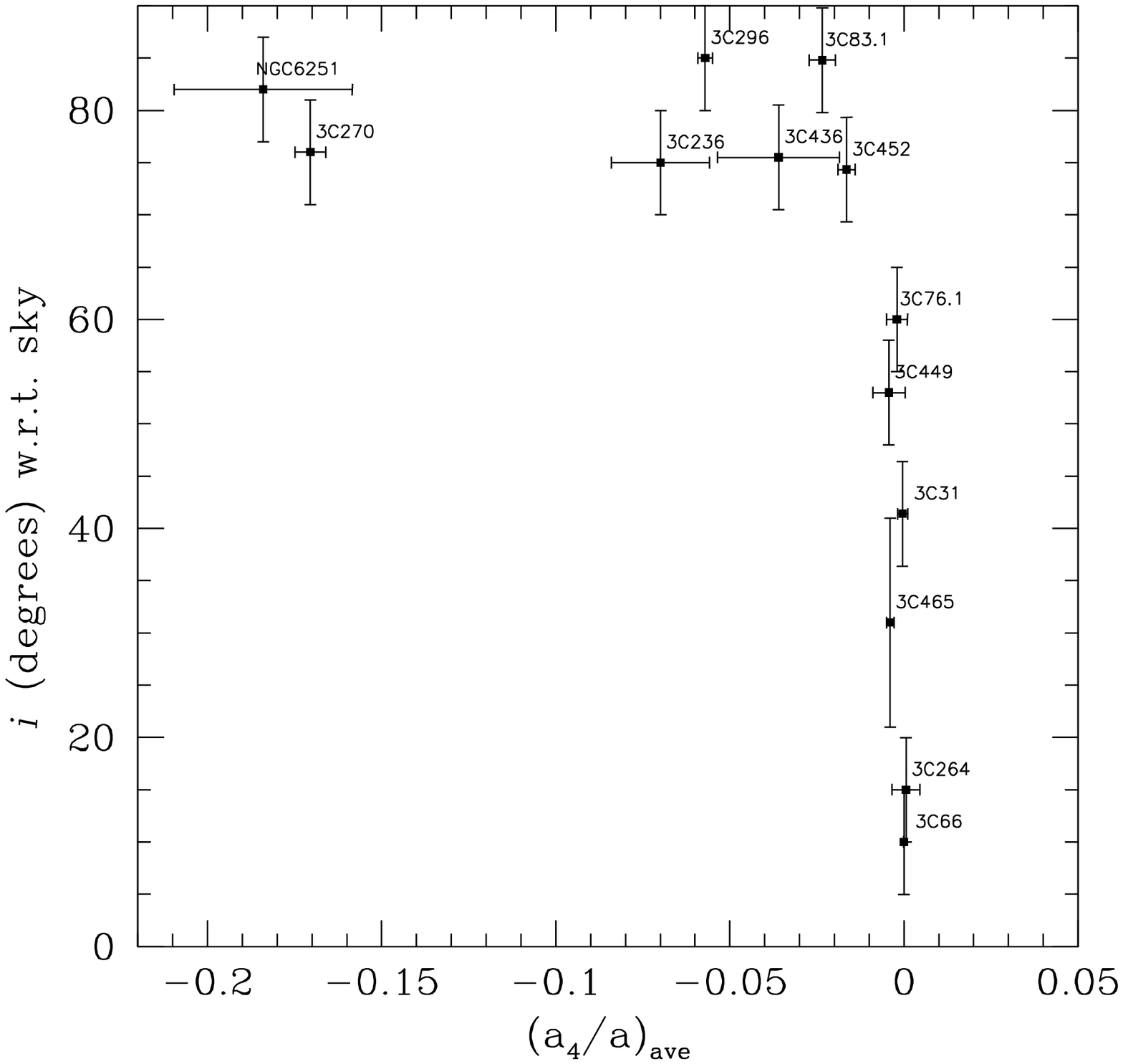}{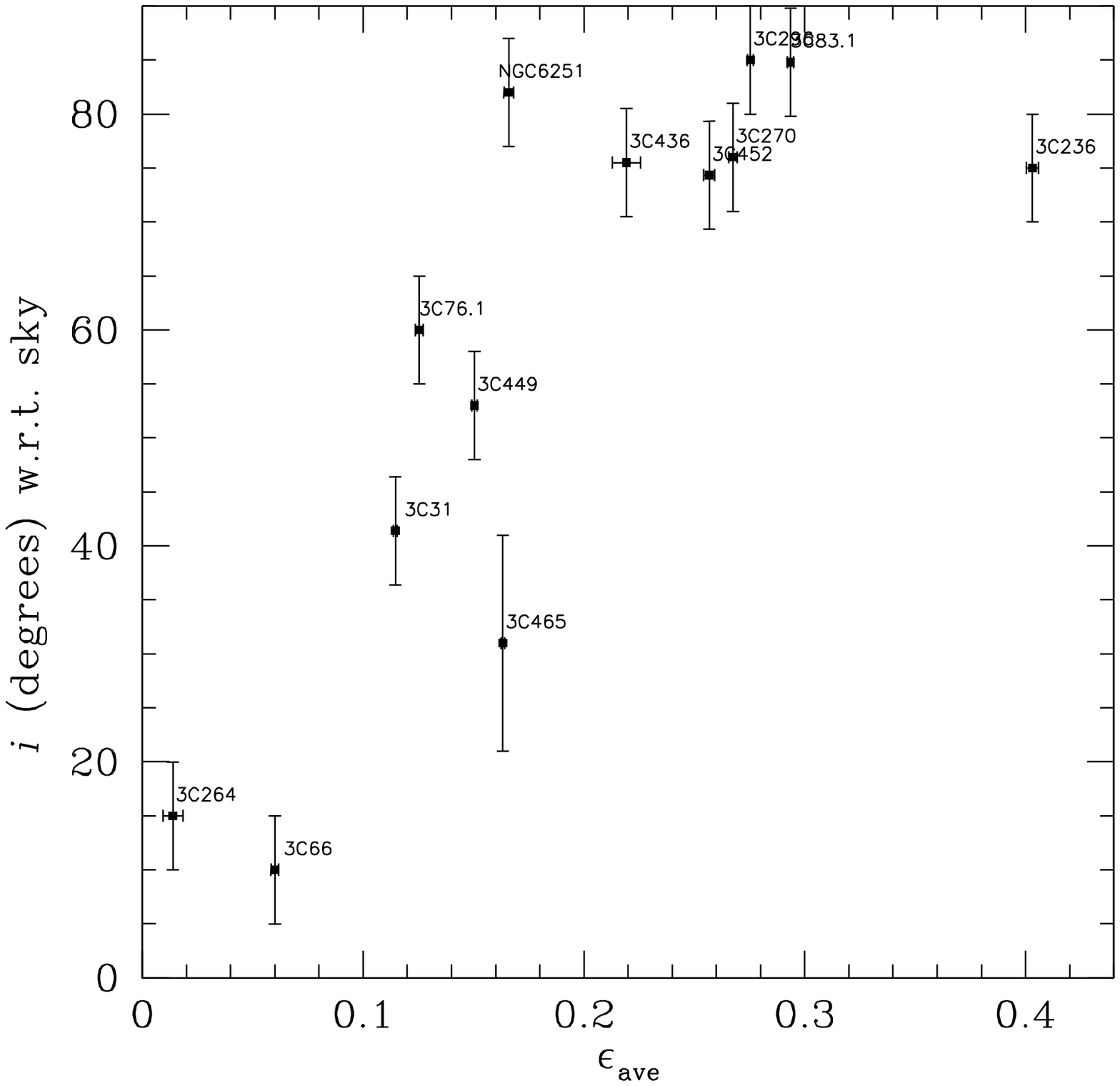}
\caption{({\it a}) Inclination $i$ (in degrees and with respect to the
plane  of  the sky)  of  the  13 dusty  disks  in  our sample  vs.~the
$\left(a_4/a\right)_{\mathrm{ave}}$  ``boxiness''  parameter of  their
host galaxy isophotes.  Dusty  disk inclinations $i$ are estimated for
each disk by $i= \arccos b/a$,  where $b/a$ is the ratio of the disk's
minor  and  major  axis lengths.   $\left(a_4/a\right)_{\mathrm{ave}}$
values are  calculated from a region-specific average  of $a_4/a$, the
$\cos 4\alpha$  Fourier coefficient normalized to the  semi major axis
of the isophote  at which it was measured  (see \S2).  Negative values
of  $\left(a_4/a\right)_{\mathrm{ave}}$  are  indicative  of  ``boxy''
isophotes,  whereas  positive values  are  manifestation of  ``disky''
isophotes.   Note   the  correlation  between   disk  inclination  and
isophotal boxiness.  ({\it b})  Dusty disk inclination $i$ vs.~average
isophotal ellipticity $\epsilon_{\mathrm{ave}}$.   As was the case for
boxiness,  there  also  appears  to  be a  relationship  between  disk
inclination and ellipticity.  }
\label{fig:trend}
\end{figure*}

In Fig.~\ref{fig:disklane}{\it  a} we present our  color-map for 3C~31,
which contains  an excellent  example of a  dusty disk.   Although the
largest  in our  sample (at  $\sim 2.5$  kpc in  diameter),  this disk
shares similar  properties with  the 12 others  we observe.   Based on
their common traits, we define a dusty disk as a round nuclear feature
seen in absorption, qualitatively appearing continuous (from its outer
boundaries to the nucleus), flat, and sufficiently resolvable so as to
allow for a confident estimate of $b/a$ (the ratio of the disk's minor
axis $b$ to its major axis $a$).  Assuming a circular, flat disk, this
axis  ratio  can  be  projected   into  an  estimate  for  the  disk's
inclination $i$ with  respect to the plane of the sky  by $i = \arccos
\left( b/a  \right)$.  These disks are typically  associated with more
than 1  $R$-band magnitude of  extinction.  In many instances,  it was
necessary to  first produce an $H/R$-band  color map so  as to confirm
the  dust feature was  indeed continuous  from its  outer edge  to the
unresolved nucleus.   Many of  the dusty disks  in our  sample possess
small asymmetries,  including lopsidedness (e.g.,  3C~465, giving rise
to    added     uncertainty    in    its     $b/a$    estimate;    see
Fig.~\ref{fig:appfig}),  non-planar  warps  (e.g.,  NGC~6251,  3C~449,
Fig.~\ref{fig:appfig},    \citealt{ferrarese99,tremblay06}),    spiral
features  (e.g.,  3C~31,  Fig.~\ref{fig:disklane}{\it a}),  and  wispy
filamentary structures (e.g., 3C~66B).  As our study is concerned with
the  orientations of  disks as  measured  with respect  to their  {\it
outer} edges,  we are generally  not concerned with  these small-scale
features.

In  Fig.~\ref{fig:disklane}{\it  b}  is  a  color-map  of  the  highly
disturbed dust  feature in 3C~321, which  we classify as  a lane.  The
three other  features defined  as lanes in  our sample are  similar in
appearance, being  ``arc''-like instead of  flat like a disk,  and are
unlikely to extend all the way  to the nucleus.  Instead, the lanes in
our sample appear to reside at much greater distances from the nucleus
than do  the disks, and the projected  extent of a lane  is always far
greater than  a disk  diameter.  For these  reasons, it is  not likely
that   we   are   misidentifying   edge-on   disks   as   lanes.    In
Fig.~\ref{fig:radhist} we present the distribution of linear sizes for
all of the disks and lanes in our sample.  Lanes are associated with a
swath of extinction (greater  than 1 $R$-band magnitude) spanning more
than  $\sim 4$  kpc,  while the  disks  in our  sample have  diameters
smaller than $\sim 2.5$ kpc.

Some  objects  in  our  sample  (e.g., 3C~317,  3C~338)  contain  thin
tendrils and  wisps of  dust, which we  do not consider  lanes.  While
lanes are  typically associated with more than  one $R$-band magnitude
of extinction, far less massive  and cohesive features such as the one
in  3C~338 are  far more  optically thin  at $R$-band  and  are better
described as ``wispy tendrils'' or  clumps rather than full lanes.  We
also  note that  3C~236 is  unique  in that  its inner  dusty disk  is
connected  to a  large surrounding  filamentary structure.  We  do not
classify this outer feature as a lane as it appears to be connected to
the  inner disk  by  a  thin tendril.   Nevertheless,  we measure  the
inclination of  the dust  feature in 3C~236  with respect to  the {\it
inner}  nuclear disk.  We discuss  3C~236  further in  \S6. See  Table
\ref{tab:tab1}  for a  qualitative classification  of  dust morphology
(where applicable) for every object in our sample.

\vspace{0.2in}          

\subsection{Host galaxy isophotal properties}

Here  we  examine  the  sample-wide  distributions  of  the  isophotal
parameters           $\left(a_4/a\right)_{\mathrm{ave}}$           and
$\epsilon_{\mathrm{ave}}$,  which  we  calculated using  the  strategy
discussed in \S3.  The  histograms for these distributions are plotted
in  Figs.~\ref{fig:isohist}{\it  a} and  {\it  b}, respectively.   The
shaded  regions of both  histograms mark  the 13  objects in  which we
observe dusty disks, the morphologies of which we describe in \S4.1.

From    the   $\left(a_4/a\right)_{\mathrm{ave}}$    distribution   in
Fig.~\ref{fig:isohist}{\it a}, it is clear that our sample favors boxy
(e.g.   $-0.1  < \left(a_4/a\right)_{\mathrm{ave}}  <  0$) over  disky
(e.g.  $0 <  a_4/a < 0.1$) objects by one-third  to one-quarter of the
sample,  respectively.  Of  greater interest  is the  fact  that dusty
disks only appear in elliptical  (or round) and boxy hosts (e.g., with
negative values  of $\left(a_4/a\right)_{\mathrm{ave}}$), and  are not
observed in disky objects (with $a_4/a > 0$) at all.  The distribution
of  mean  ellipticities  shown  in  Fig.~\ref{fig:isohist}{\it  b}  is
decidedly        flatter        than        the       spread        in
$\left(a_4/a\right)_{\mathrm{ave}}$, and is  representative of a wider
range  of intrinsic  ellipticities  observed in  active and  nonactive
ellipticals.

The most significant result here is  that dusty disks are not found in
disky galaxies. We expand upon  this finding in the following section,
in  which we examine  possible connections  between the  inclinations of
these  disks   and  the  specific  $\left(a_4/a\right)_{\mathrm{ave}}$
values of the boxy hosts in which they reside.

\subsection{A trend in dusty disk inclination and isophote boxiness}

Here  we  compare isophotal  boxiness  to  dust  morphology for  those
objects observed to contain nuclear disks.  In \S4.1, we described how
we estimate the inclinations of  these disks with respect to the plane
of  the   sky.   We  plot  these  inclinations   vs.   the  calculated
$\left(a_4/a\right)_{\mathrm{ave}}$   of   their   host  galaxies   in
Fig.~\ref{fig:trend}{\it a}.  Despite the inherent uncertainties, this
plot brings to attention several interesting properties.

As was  the case  for Fig.~\ref{fig:isohist}{\it a},  we see  again in
Fig.~\ref{fig:trend}{\it a}  that no circumnuclear  disks are observed
in  disky  host  galaxies.   More  specifically,  these  disks  reside
exclusively  in host galaxies  with $\left(a_4/a\right)_{\mathrm{ave}}
<0$, ranging from round (i.e., 3C~264) or elliptical (i.e., 3C 449) to
boxy (i.e., 3C~296) and very boxy (i.e., NGC~6251).

Moreover, the distribution of these objects appears to follow a trend.
While face-on or moderately inclined disks are all clustered very near
$\left(a_4/a\right)_{\mathrm{ave}} \approx 0$  with very little spread,
nearly  edge-on  disks  are  seen  only in  clearly  boxy  hosts  with
$\left(a_4/a\right)_{\mathrm{ave}}$   ranging   nearly   an  order   of
magnitude in negative values.

Comparing this result to  Fig.~\ref{fig:trend}{\it b} in which we plot
$\epsilon_{\mathrm{ave}}$ against $\left(a_4/a\right)_{\mathrm{ave}}$,
it  is apparent  that, at  least for  the low  to  moderately inclined
disks, ellipticity  roughly correlates  to inclination.  That  is, low
inclination disks  (e.g., 3C~66B, 3C~264, and 3C~76.1)  tend to reside
in {\it  round} host galaxies  with $\left(a_4/a\right)_{\mathrm{ave}}
\approx \epsilon_{\mathrm{ave}} \approx 0$.  Moderately inclined disks
(e.g., 3C~465, 3C~31, and 3C ~449) appear to reside in hosts with {\it
elliptical}    isophotes   having   $\left(a_4/a\right)_{\mathrm{ave}}
\approx    0$    and    an    intermediate   or    high    value    of
$\epsilon_{\mathrm{ave}}$.  On  the other hand,  there is considerably
higher  spread in $\epsilon_{\mathrm{ave}}$  for the  highly inclined,
boxy galaxies.  NGC~6251, for example, is by far the boxiest object in
Fig.~\ref{fig:trend}{\it a},  but is seen  in Fig.~\ref{fig:trend}{\it
b} to exhibit only moderate ellipticity.

To restate  the main  result of this  analysis: we find  that galaxies
with dusty disks exhibit a correlation between their average isophotal
boxiness  and  the  inclination  of  the  disk  with  respect  to  the
sky. Face-on disks are observed  in round galaxies while edge-on disks
are  seen   only  in  boxy   galaxies.   Furthermore,  we   observe  a
relationship between disk inclination  and ellipticity that holds more
strongly for objects with low or moderately inclined disks than it does
for those with  edge-on disks.  In the following, we  will make use of
the latter result to explore the connections between the properties of
the galaxies that possess lanes and their orientation.

\subsection{Estimating orientations of objects with dust lanes}

\begin{figure*}
\plottwo{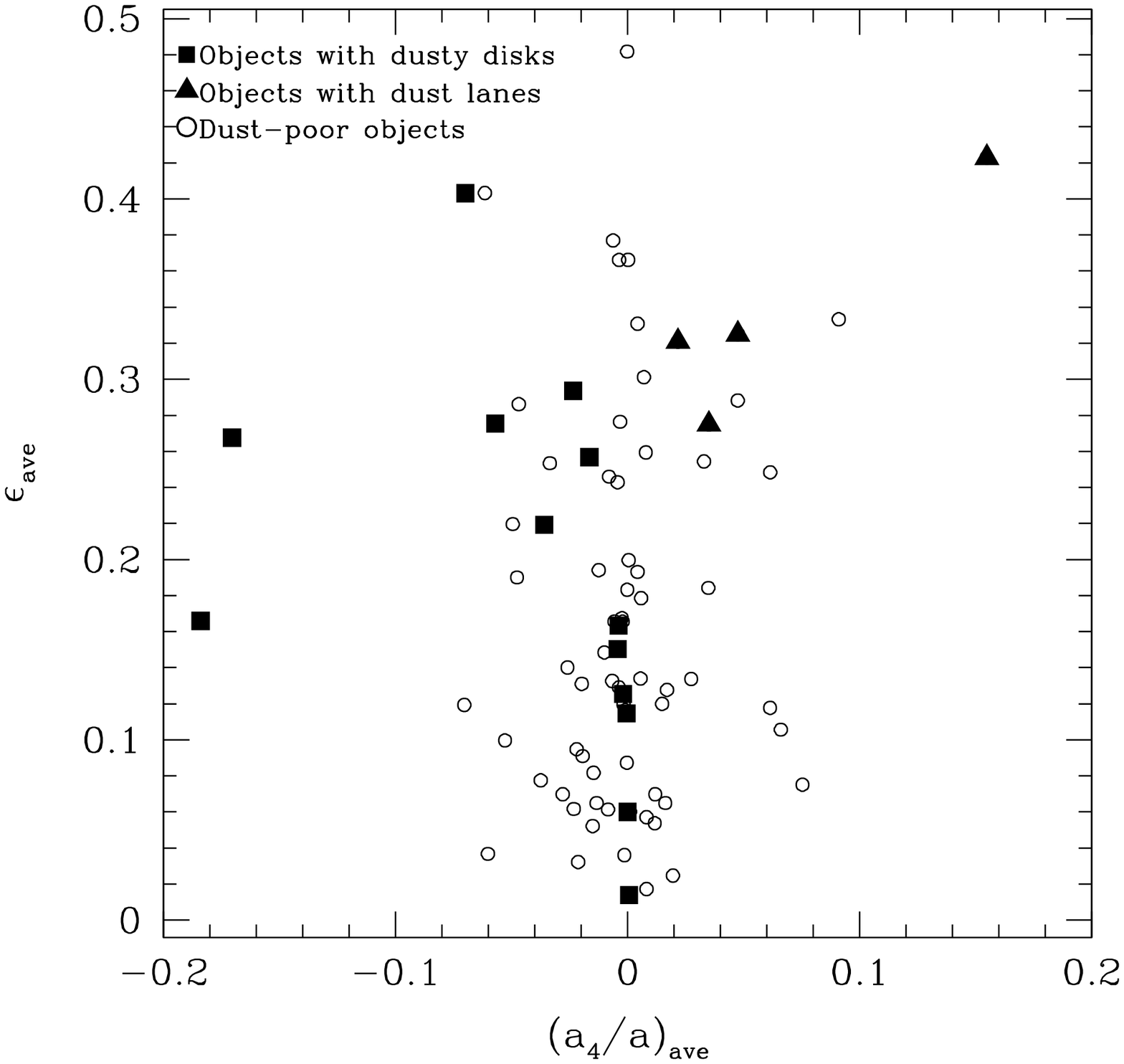}{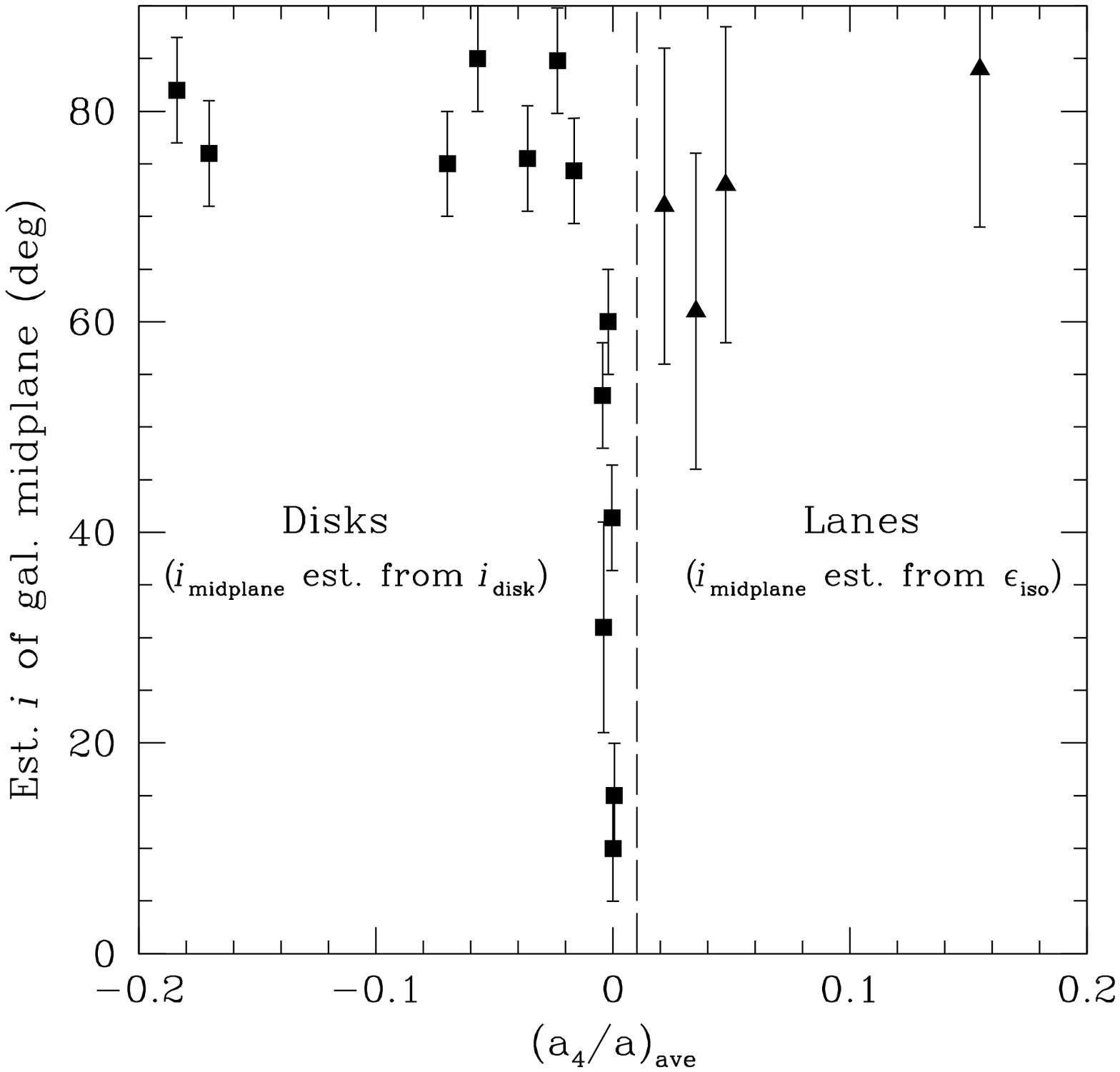}
\caption{  ({\it  a})  Average  ellipticity  $\epsilon_{\mathrm{ave}}$
vs.~average isophotal boxiness $\left(a_4/a\right)_{\mathrm{ave}}$ for
each object in  our sample.  ({\it b}) A  {\it qualitative} estimation
of the  inclination $i_{\mathrm{midplane}}$ (in  degrees, with respect
to the  plane of the sky)  of the galactic midplanes  of the dust-rich
host galaxies  in our sample.   For hosts containing disks  (which are
expected   to    reside   in   a   symmetry    plane),   we   estimate
$i_{\mathrm{midplane}}$  directly  by  disk  inclination.   For  hosts
containing  lanes, we  estimate $i_{\mathrm{midplane}}$  based  on the
loose   ellipticity-disk    inclination   relationship   apparent   in
Fig.~\ref{fig:trend}{\it  b}  (and  discussed  in  \S5),  as  well  as
qualitative  appearance.  The  figure  is intended  to illustrate  the
apparent   bimodality   in   isophotal   structures   of   disk-   and
lane-containing hosts.}
\label{fig:a4ellip}
\end{figure*}

To  provide sample-wide context,  in Fig.~\ref{fig:a4ellip}{\it  a} we
plot                  $\epsilon_{\mathrm{ave}}$                 versus
$\left(a_4/a\right)_{\mathrm{ave}}$  for  the  whole  sample.   Filled
squares  represent  objects with  dusty  disks,  filled triangles  are
objects with dust lanes, and empty circles are ``dust poor'' galaxies.
Despite   the    fact   that   objects   with    extreme   values   of
$\left|\left(a_4/a\right)_{\mathrm{ave}}\right|$   are   present,  our
sample    is    clearly    clustered    around   small    values    of
$\left|\left(a_4/a\right)_{\mathrm{ave}}\right|$.   However, we stress
that         even         a         small        difference         in
$\left|\left(a_4/a\right)_{\mathrm{ave}}\right|$  between two galaxies
implies a significant difference  in the structure of their respective
isophotes.   As  we  move  to  bins of  greater  degrees  of  boxiness
(leftward   on   the   $\left(a_4/a\right)_{\mathrm{ave}}$   axis   of
Fig.~\ref{fig:a4ellip}{\it  a}),  the  fraction  of bins  occupied  by
objects  with  dusty disks  roughly  increases proportionately.   While
roughly      25\%      of       the      bin      centered      around
$\left(a_4/a\right)_{\mathrm{ave}} =  0$ consists of  hosts with dusty
disks, 100\% of the boxiest bins contain disks.
Note also that the four objects  with lanes are all located in the top
right-hand-corner of the  plot.  This means that lanes  are present in
galaxies with both disky and  high ellipticity isophotes.  
In \S4.3 we  found a trend between ellipticity  and orientation of the
dusty disks.   Since those  disks are closely  aligned with  the major
axis  of the  galaxies  in  which they  reside  \citep{tran01}, it  is
tempting to derive an ``orientation'' for the galaxies with lanes.

In
Fig.~\ref{fig:a4ellip}{\it  b} we  present a  plot to  extend  that of
Fig.~\ref{fig:trend}{\it a}  to include the  lanes, for which  we plot
large error bars due to the uncertainty in the inclination-ellipticity
relation.   While on  the left  side  of the  plot both  high and  low
inclination  disks  are  present,  on  the  right  side  we  lack  low
ellipticity galaxies with lanes. With only 4 galaxies in the sample it
is difficult to establish whether this  is due to a selection bias or
it is a real effect. Regardless, this figure serves to again illustrate 
that lanes are observed only in disky hosts, as well as suggests 
at a possible dependence of diskiness on midplane inclination, just as we observe for 
our larger sample of dusty disks in boxy galaxies.

\subsection{Comparison of isophotes with radio luminosities and core-dominance ratios}

\begin{figure*}
\plottwo{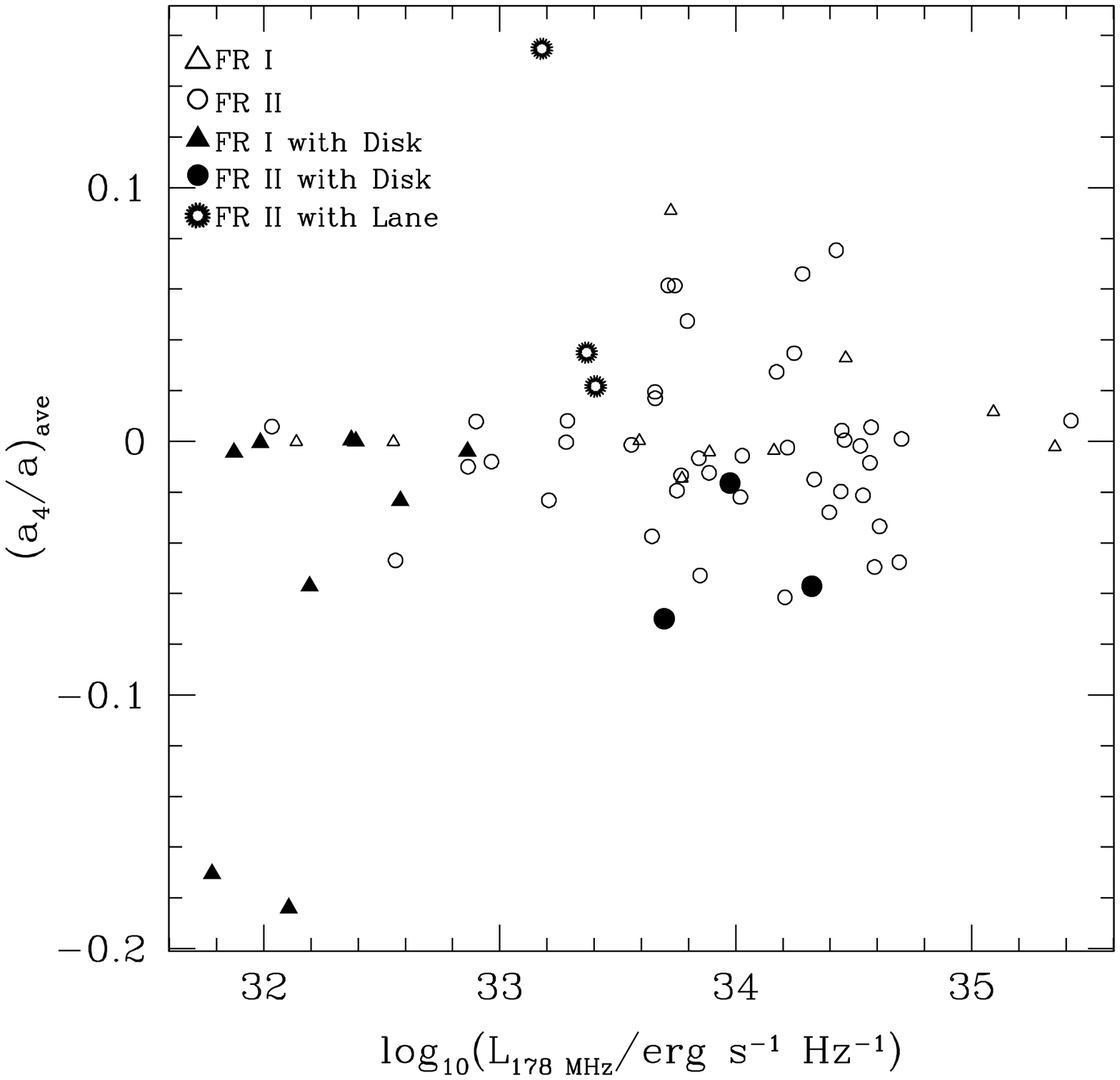}{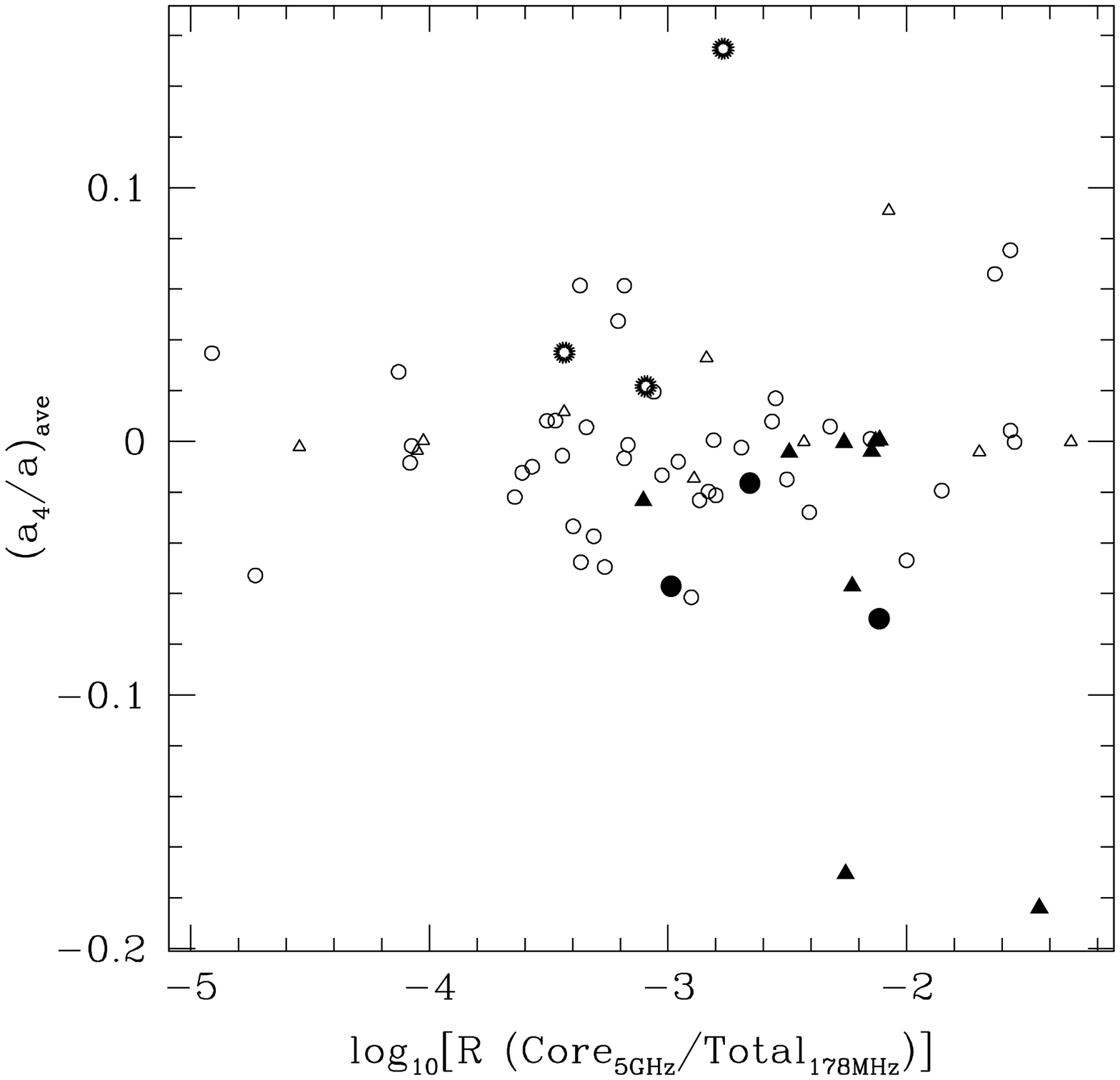}
\caption{  ({\it a})  Average isophotal  boxiness vs.~the  log  of the
total radio luminosity at 178  MHz (values taken from NED).  ({\it b})
The  same boxiness  distribution  in ({\it  a})  vs.~5 GHz  / 178  MHz
core-dominance  ratios $R$.   More core-dominated  galaxies  (like NGC
6251, the lowest  black triangle in the plot) are  closer to the right
edge of the figure.  }
\label{fig:radio}
\end{figure*}

In Fig.~\ref{fig:radio}{\it a} we plot the sample-wide distribution of
total  radio luminosities  at 178  MHz (retrieved  from  the NASA/IPAC
Extragalacitc  Database)  against $\left(a_4/a\right)_{\mathrm{ave}}$,
so  as  to explore  possible  connections  between radio-loudness  and
isophotal structure.  We observe no  apparent trend as there is fairly
heavy scatter, though we note  that FR Is with dusty disks exclusively
occupy the low-power end (left-hand side) of the plot, while dust-poor
FR Is in are spread across the whole range.

Galaxies whose radio emission is dominated  by the core at 5 GHz imply
a measure of beaming of the radio jet, while observations dominated by
emission at  178 MHz suggests  that the viewer is  observing unbeamed,
more edge-on jet lobes.  The  core-dominance ratio $R$ of core (5 GHz)
to extended region  (178 MHz) fluxes is therefore  a rough orientation
indicator of  the line of sight  with respect to the  central AGN.  We
plot        $\left(a_4/a\right)_{\mathrm{ave}}$       vs.~$R$       in
Fig.~\ref{fig:radio}{\it b} by  division of core flux at  5 GHz by the
total 178 MHz  flux.  Again, we observe no trend  in this figure given
the scatter.  We do note that NGC 6251, the bottom-most black triangle
in Fig.~\ref{fig:radio}{\it  b}, is highly core-dominated,  but is the
boxiest object in  our sample with a nearly edge-on  dusty disk in its
nucleus.  While  this would  normally be surprising,  the disk  in NGC
6251 has been modeled with a  warp, such that the outer regions of the
disk  are edge-on and  aligned with  the isophotal  major axis  of the
galaxy, while the inner accretion region is oriented such that the jet
is roughly  pointed toward the  observer, accounting for  the apparent
discrepancy (e.g.,~\citealt{ferrarese99, chiaberge03}).

In both Figs.~\ref{fig:radio}{\it  a} and {\it b} we  use triangles to
denote  FR I  galaxies and  circles to  denote FR  II  galaxies. Empty
symbols indicate  that the object is ``dust-poor'',  and black symbols
mark the presence  of a dusty disk. Bold empty  circles indicate an FR
II in  which we observe a  lane. We do not  observe any lanes  in FR I
galaxies.  We  also note that,  while powerful, edge-brightened  FR II
galaxies outnumber the low-power, edge-darkened FR Is in our sample by
over 2:1, we find  that dusty disks reside in over twice  as many FR I
galaxies as in FR IIs.  Moreover,  all of the dust lanes in our sample
are found  in FR II  hosts.  This is  consistent with the  findings of
\citet{dekoff00}, who  noted that dust  in FR I galaxies  is typically
found in the  form of a nuclear  disk below 2.5 kpc in  size, while FR
IIs more commonly possess dust in clumpy patches and filaments.  While
these  results  are  interesting,  we  must be  weary  of  a  possible
observational bias,  as FR II  galaxies are typically found  at higher
redshifts than FR Is.

\section{Implications of our Results}

In \S4.3 we describe an apparent relationship between the inclinations
of circumnuclear dusty  disks in our sample and  the boxiness of their
host galaxy isophotes.  We observe  that edge-on dusty disks reside in
boxy  host  galaxies, while  face-on  disks  are  seen only  in  round
galaxies.   Intermediately inclined  disks reside  in  more elliptical
galaxies.  Dust lanes, on the other hand, are only found in hosts with
moderate to high levels of diskiness.  Dusty disks reside in 13 of the
33 objects  with dust observed  in varying distributions  through {\it
HST}/WFPC2 optical  imaging. These 33 objects  are a subset  of our 84
object sample  with {\it HST}/NIC2 near-infrared  imaging.  Our sample
of hosts  possessing dusty  disks is free  from orientation  bias (see
\S2) and we find no exceptions
to   the   apparent   inclination-boxiness  trend   described   above.
\citet{martel00} noticed a  similar trend in their optical  study of 7
of the dusty  disks we describe here.  In our  sample, nearly twice as
large as that  of \citet{martel00}, the trend has  only grown stronger
and more convincing.  In this section, we argue that our results carry
interesting  implications   with  regards  to   the  manifestation  of
isophotal structure, as well as the post-merger history of dust in the
settling sequences of boxy and disky hosts.

\subsection{Is boxiness dependent on viewing angle?}

Here we discuss whether or not our results suggest that boxiness is an
orientation-dependent     effect.       It     is     evident     from
Figs.~\ref{fig:trend} and \ref{fig:a4ellip}  that nearly edge-on dusty
disks  reside in  boxy galaxies  that span  an order  of  magnitude in
values   of   $\left(a_4/a\right)_{\mathrm{ave}}$.   Conversely,   the
moderate- to low-inclination disks are seen only in host galaxies with
very  similar values  of  $\left(a_4/a\right)_{\mathrm{ave}}$, all  of
which are  nearly zero  (though all are  just slightly  negative). The
fact  that such  tight clustering  is absent  for the  highly inclined
disks  may  reflect  intrinsic  differences in  the  stellar  velocity
dispersions  of  these  galaxies,   as  varying  degrees  of  velocity
anisotropy  and orbital  triaxiality  are thought  to produce  varying
degrees of boxiness (e.g., \citealt{bender88,hao06}).

We may associate the inclinations of dusty disks with the inclinations
of the  galactic midplanes in which  they are expected  to settle. The
rotation axis of the disk is therefore aligned with a symmetry axis of
the  underlying galaxy.   Keeping this  in mind  and  considering both
Figs.~\ref{fig:trend}{\it  a}  and  {\it  b} together,  a  qualitative
picture of the possible dependence of boxiness on viewing angle begins
to  emerge.   Were we  to  view a  test  galaxy  with a  line-of-sight
coincident  to its  symmetry  axis, we  would  view its  perpendicular
midplane as  face-on.  As  the inclination of  the symmetry  axis with
respect  to  the  line-of-sight  increases to  more  moderate  levels,
Figs.~\ref{fig:trend}{\it  a}  and  {\it  b}  suggest  that  perceived
boxiness would remain  nearly nonexistent, while isophotal ellipticity
would increase  significantly with the change in  viewing angle.  Only
when we are viewing the galaxy's  symmetry axis nearly in the plane of
the  sky (such  that the  galactic midplane  is edge-on)  may boxiness
become  fully apparent  in the  isophotes.   For example,  if we  were
somehow  able to  view  the face-on  disk  in 3C~66B  as edge-on,  the
otherwise round isophotes of its host might appear boxy.

Our sample is randomly oriented, and there is no reason to expect that
face-on  disks and edge-on  disks reside  in different  populations of
galaxies.  We therefore  conclude that boxiness is, at  least in part,
an  orientation-dependent  phenomenon.  This  is  consistent with  the
results of  early $N$-boxy simulations of galaxy  mergers, which found
that the  boxiness of a remnant  is dependent on  viewing angle (e.g.,
\citealt{limaneto95}).   It  is  tempting  to  surmise  that  such  an
orientation dependence extends to  disky isophotes, as well.  The four
dust lanes in our sample are found only in disky galaxies, though this
small number  prevents us from drawing conclusions  from their already
uncertain   distribution  in   Fig.~\ref{fig:a4ellip}{\it   b}.   Past
simulations of  mergers that account for gas  dissipation suggest that
our  results have  relevance with  regards to  the role  of dust  as a
tracer of merger history, which we discuss this below.

\subsection{Dust as a tracer of merger history}

Here  we qualitatively discuss  the implications  of our  results with
respect to the role of gas dynamics in the formation of boxy and disky
merger remnants.   It is clear  that there is an  intrinsic connection
between  the morphology  and  inclination of  dust  and the  isophotal
structure of  the early type host  in which it resides.   Hardly a new
idea, such a  connection is expected on at  least a theoretical level.
Gas and  dust acquired through a cannibalistic  merger should coalesce
on a few  dynamical timescales and precess in a  symmetry plane of the
newly formed remnant, finally settling  into the potential well of the
host           on           a           precession           timescale
(e.g.,~\citealt{gunn79,tubbs80,tohline82,habe85}).  All of this should
occur on a timescale of order a Gyr, during which time the dust should
dissipate angular momentum  at a rate dependent upon  the structure of
the  potential well  and the  star  formation efficiency  of the  gas
\citep{barnes96,bekki97}.     In     the    scenario    proposed    by
\citet{lauer05}, filamentary  distributions of dust that  have not yet
reached the nucleus would be  classified as dust lanes, which might be
thought  of  as transient  structures  that  would  eventually form  a
nuclear disk if given sufficient time.

Dynamically,  the  presence of  added  gas  and  dust in  the  central
potential  is  thought  to  alter  the  shape  of  the  well,  causing
destabilization in  central stellar box orbits and  thus affecting the
structure of  the host  galaxy's isophotes \citep{barnes96}.   This is
supported by numerous simulations of dissipational mergers, which find
that higher gas/dust content in one or both of the progenitor galaxies
typically  leads to more  pronounced diskiness  in the  remnant (e.g.,
\citealt{springel00},  and references  therein).  On  the  other hand,
boxy galaxies are formed  through gas-poor mergers of progenitors with
high-density    stellar    bulges    of    relatively    equal    mass
(e.g.,\citealt{naab03,khochfar05}).

Our results are consistent with  the notion that dust acquired through
a  merger has  a direct  impact on  the manifestation  of  boxiness or
diskiness in  the remnant.  We  find, however, that the  nuclear dusty
disks  in our  sample  are exclusively  associated  with either  boxy,
elliptical, or  round hosts,  and are never  observed in  disky hosts,
which are seen only to contain dust lanes.  This seems contrary to the
idea that  dust-rich (``wet'') mergers produce disky  remnants {\it if
we hold  that disks and lanes  are both external in  origin}.  In that
case,  the  presence of  a  disk indicates  that  a  large portion  of
externally acquired  dust has  been allowed to  settle in  the nucleus
following  the  merger,  thereby  disrupting stellar  box  orbits  and
washing out isophotal boxiness (according to theory and simulations).

We observe  the opposite  to be true,  finding that  the circumnuclear
dusty disks in our sample  exclusively prefer boxy hosts.  Recall that
the  3CR sample should  be free  from orientation  bias, and  that the
total  $\left(a_4/a\right)_{\mathrm{ave}}$   distribution  of  our  84
targets is largely isotropic, (e.g.,  while the majority of our sample
is boxy, nearly  one-third of our targets would  qualify as disky; see
Fig.~\ref{fig:isohist}).    Were  there   not  {\it   some}  intrinsic
preference for boxiness among the population containing nuclear disks,
it is statistically  likely that we would have found  dusty disks in a
few disky  hosts.  Instead, dust-rich disky ellipticals  in our sample
only possess dust in the form of lanes.

Our results might  be reconciled with those of  past simulations if we
consider  a scenario  in which  dusty  disks and  lanes have  distinct
origins.   This  has been  suggested  before,  based  on the  observed
tendency for  dusty disks, not lanes,  to align with  the isophotes of
their   hosts   \citep{tran01}.   That   study   offered  a   possible
explanation, which our results support:
\begin{itemize}
\item  Dusty disks are  {\it native}  to the  nucleus of  a progenitor
galaxy  with a  high-density stellar  core.  In  an  equal-mass merger
destined to produce a boxy remnant (as predicted by simulations), this
dust remains  in the nucleus of  the parent, and  any perturbations it
experiences vanish rapidly. In  this scenario, dusty disks are ancient
structures, and are comprised of  dust that has resided in the nucleus
of a massive progenitor {\it prior} to the last merger event.
\item Filamentary dust  lanes are truly {\it external}  in origin, and
are the product  of tidal stripping or the  recent collision of small,
unequal mass, gas-rich progenitors with significant dust mass residing
in the  outer extents of their  volumes.  In the  disky remnant, rapid
stellar rotation and an ISM  recently shocked by the collision prevent
the   newly    acquired   dust   from    efficiently   losing   angular
momentum. Instead of settling into the nucleus, it is distributed in a
filamentary fashion  at more distant radii.  This  would be consistent
with findings that lanes do not typically align with the major axis of
their host, while the opposite is true for dusty disks.
\end{itemize}

Our  results  are  consistent   with  those  of  previous  studies  in
suggesting that the elliptical hosts  of nuclear dusty disks and lanes
are  drawn  from  two   different  galaxy  populations  with  distinct
formation histories.  We conclude that disks reside in old remnants of
gas-free  ``dry''  mergers,  while  lanes primarily  reside  in  young
remnants of gas-rich ``wet''  mergers.  This comprises the second main
conclusion  of  this  paper.

\section{Summary and Discussion}

In  this  paper we  have  examined  possible  connections between  the
morphology of  dust distributions in radio galaxies  and the isophotal
properties of  their elliptical  hosts.  We obtain  deep near-infrared
{\it HST}/NIC2 and {\it TNG} imaging of a sample of 84 radio galaxies,
the majority  of which are found  in the 3CR  catalog of extragalactic
sources.  For a 33 object subset of our sample identified as dust-rich
by  previous {\it  HST} optical  studies, we  retrieve  companion {\it
HST}/WFPC2 $R$-band imaging  of the nuclear regions of  each host.  We
create $H$/$R$-band NIC2/WFPC2 color-maps for each dusty host so as to
more clearly  resolve the dust  features contained within.   Among our
dust-rich  subset, we  identify  13 circumnuclear  dusty  disks and  4
filamentary lanes, while the remaining dust-rich hosts possess dust in
less  settled  distributions  like   clumps  and  patches.   From  the
luminosity profiles  extracted from  the near-infrared images  of each
galaxy  in our  sample, we  qualitatively select  radial  regions over
which    we    calculate     the    isophotal    structural    metrics
$\left(a_4/a\right)_{\mathrm{ave}}$  (the   average  Fourier  boxiness
coefficient) and $\epsilon_{\mathrm{ave}}$ (the average ellipticity of
the isophotes).

This  allows for  a comparison  between the  distribution of  the dust
features in our  sample and the isophotal boxiness  and ellipticity of
the hosts  in which  they are found.   Among those  objects containing
dusty disks and lanes, we find the following:
\begin{itemize}
\item Edge-on (with respect to the line of sight) dusty disks are seen
only    in    host     galaxies    with    boxy    isophotes    (e.g.,
$\left(a_4/a\right)_{\mathrm{ave}} <  0$; see Fig.~\ref{fig:trend}{\it
a}).
\item Face-on dusty  disks are seen only in  host galaxies with nearly
round  isophotes (e.g., $\left(a_4/a\right)_{\mathrm{ave}}  \approx 0$,
low $\epsilon_{\mathrm{ave}}$, see Figs.~\ref{fig:trend}{\it a} and {\it b}).
\item Intermediately-inclined  dusty disks  are seen in  host galaxies
with  elliptical  isophotes (e.g.,  $\left(a_4/a\right)_{\mathrm{ave}}
\approx     0$,     intermediate    $\epsilon_{\mathrm{ave}}$,     see
Figs.~\ref{fig:trend}{\it a} and {\it b}).
\item  Dust  lanes  are   seen  exclusively  in  disky  host  galaxies
(e.g. with  pointed isophotes, $\left(a_4/a\right)_{\mathrm{ave}}  > 0$).  
\item No  disky host  galaxies are observed  to contain  dusty nuclear
disks.
\end{itemize}

We discuss  these results in  the context of  the role of dust  in the
formation of boxy  and disky merger remnants.  While  our sample of 13
dusty disks  is relatively  small, we argue  that the  connection they
share with the isophotes of  their hosts carry real implications as to
the expression of isophotal structure and the post-merger histories of
dusty disks and lanes.  In \S5, we discuss how our results corroborate
those of previous studies in suggesting that:
\begin{enumerate}
\item Isophotal  boxiness is dependent on viewing  angle.  When viewed
along the galactic axis of symmetry, the dusty disks in our sample are
face-on and  the isophotes of  the host are  round.  When the  axis of
symmetry  is highly inclined  with respect  to the  line-of-sight, the
disks are  nearly edge-on  and the isophotes  appear boxy. We  have no
reason to  suspect that the high  and low inclination  dusty disks are
residing in different populations of galaxies. The degree of perceived
boxiness in this galaxy population (e.g., those that host dusty disks)
must therefore be dependent on viewing angle.
\item Dusty disks may not be external in origin as previously thought.
Instead,  they  may be  long-time  residents  of  old gas-free  merger
remnants with high degrees of stellar velocity anisotropy accounting 
for intrinsic (though not necessarily observed) boxiness.
\item Filamentary dust lanes might share an entirely distinct history,
having been  acquired through  recent gas-rich dissipative  mergers of
low-mass progenitors. The higher  degree of isotropic stellar rotation
that  gives rise  to  the  disky isophotes  of  this population  might
account for the fact that we only  see lanes in disky hosts. Dust in a
rapidly  rotating  system  is  less  able  to  lose  angular  momentum
effectively, greatly lengthening its lifetime as a filamentary lane.
\end{enumerate}

The study of dust and  host galaxy isophotal properties helps to place
constraints on the dynamics of early-type stellar components and their
associated  potentials.  By extension,  these may  contain information
relating to the angular momentum of the last major merger, which might
play a dominant role in  the carving of galactic equipotentials and in
setting the spin axis of the central black hole.  In this regard, dust
might act as an observable gauge of otherwise inobservable connections
between AGN and the properties of their host galaxies.

\begin{figure}
\plotone{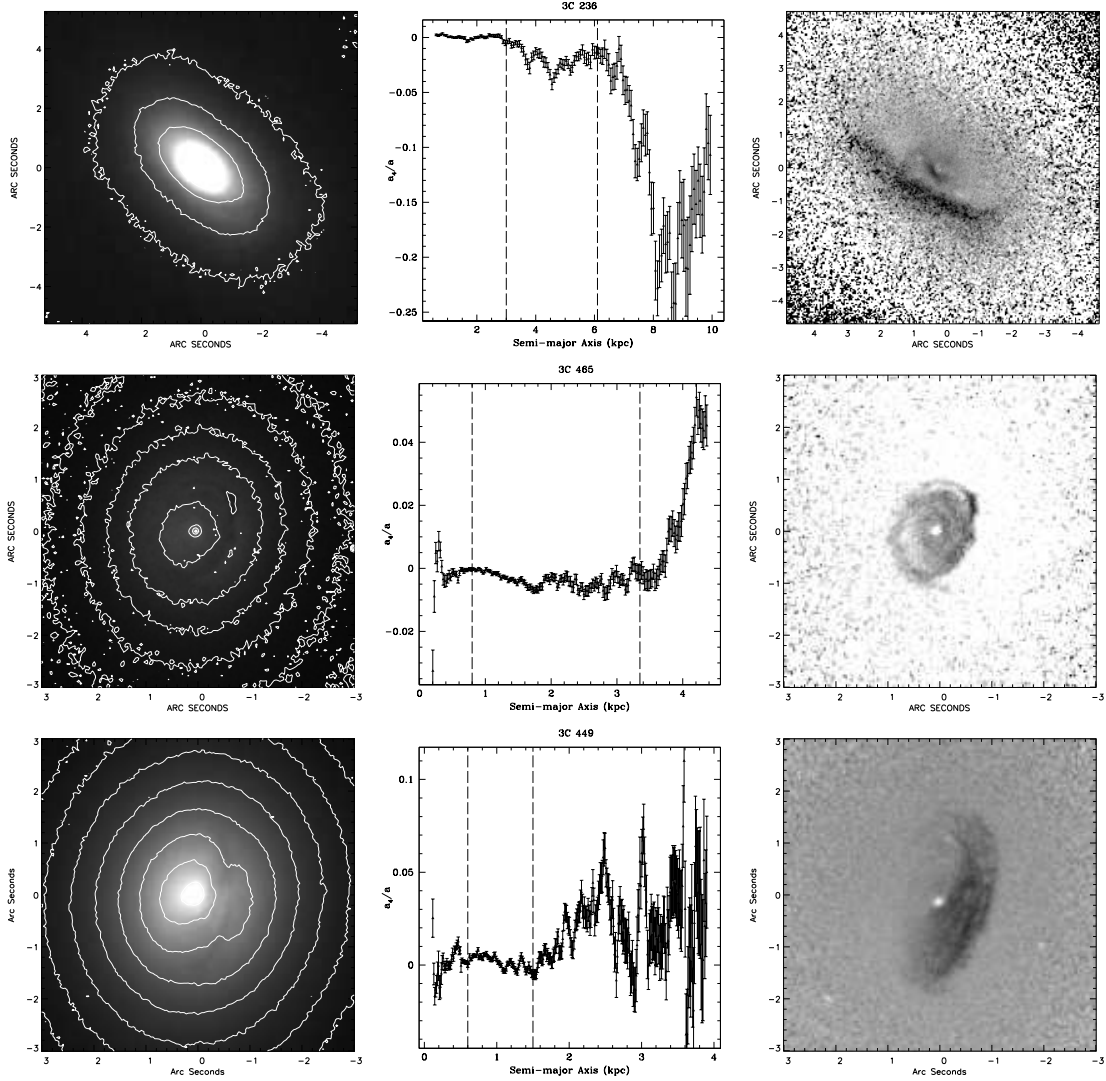}
\caption{ Examples of some of  the unique dust features in our sample,
which we  discuss directly in  \S5.  ({\it Top}) {\it  HST}/NIC2 $1.6$
$\mu$m  image  of  3C~236  with highlighted  isophotal  contours,  its
associated $a_4/a$ radial profile  with the averaging region (see \S3)
bounded by  dashed lines, and a  $1.6$ $\mu$ / 0.7  $\mu$m colormap of
the dusty  disk in the  nucleus of 3C~236,  made via division  of {\it
HST}/NIC2 and WFPC2 data.  This object  is unique in that it is one of
the largest galaxies in the  universe, and its nucleus contains both a
dusty disk and  a dust lane.  ({\it Middle}) Same  as ({\it Top}), but
for  3C~465. The disk  in 3C~465  is lopsided,  with two  sharp spiral
features to the north and  south.  ({\it Bottom}) Same as ({\it Top}),
but for 3C~449.  The dusty disk in this object is unique in that it is
a  clear outlier  to  the  observed (but  perhaps  not real)  jet/disk
orthogonality ``trend'', which we discuss  in \S1.  The disk in 3C~449
has been  modeled by \citet{tremblay06} with a  large warp, accounting
for jet/disk orthogonality on smaller ($\sim 50$ pc) scales.  }
\label{fig:appfig}
\end{figure}

The results of this paper illustrate such a possible relationship, and
even  more examples  are found  within  our own  sample.  3C~236,  for
instance, is unique  in that it contains both an  inner dusty disk and
an  outer dust  filament only  partially connected  to the  inner dust
structure   (Fig.~\ref{fig:appfig}).    \citet{odea01}  explored   the
relationship  between  the  radio  activity  of 3C~236  and  its  dust
morphology.  That  study suggested that  the AGN fuel supply  had been
cut off at  some time in the past, and AGN  activity has only recently
resumed,  accounting for  its relatively  young radio  morphology. The
isophotes  of  3C~236, while  boxy,  are  also  highly elliptical  and
irregularly shaped, suggesting that it has recently undergone a merger
event, a notion supported by its disturbed outer dust features.

Another example  of dust serving to illustrate  connections between an
AGN and  its host galaxy  is found in  3C~449, a clear outlier  to the
observed   (though  not   necessarily  real)   jet/disk  orthogonality
``trend''  discussed in  \S1 (see  Fig.~\ref{fig:appfig}).   The outer
regions of the dusty disk are aligned with the isophotal major axis of
the host, which makes a sharply acute angle with the radio jet axis on
the plane of  the sky.  The recent work  by \citet{tremblay06} modeled
the   dusty  disk  with   a  large   warp,  accounting   for  jet/disk
orthogonality on smaller ($\sim  50$ pc) scales.  That study discussed
physical mechanisms capable of creating and/or sustaining such a warp,
concluding  that its  source  most likely  comes  from a  perturbative
interaction   between   the   dusty   disk   and   highly   elongated,
anisotropically  distributed  isobars of  X-ray  emitting  gas in  the
ambient medium.  Were such anisotropy  coupled to the launching of the
jet through a  feedback interaction from the AGN, it  may play a large
role in a process  responsible for preferentially aligning dusty disks
to radio jets, accounting for jet/disk orthogonality.

Indeed, recent  observations have  revealed the interstellar  media in
the  central  few hundred  pc  of many  radio  galaxies  to be  highly
turbulent and  capable of  perturbing equilibrium morphologies  of the
dust   in   these   areas  (e.g.,~\citealt{   fabian03,   noelstorr03,
vernaleo06}).  The study of dust in relation to AGN feedback processes
might help account for observed non-equilibrium dust morphologies like
tendrils and warps that should  otherwise be short-lived. In a hot ISM
lacking a  cold counterpart, errosive processes might  even be capable
of destroying the  dust altogether through sputtering (e.g.,~\citealt{
canizares87, dwek92}).  Future  studies will further our understanding
of  the morphological  evolution  of  dust, as  well  as the  ultimate
longevity  of this  dust  in highly  energetic  environments like  the
hearts of radio galaxies.

To that  end, it is necessary that  we expand on this  work to include
larger  samples of  radio galaxies  as well  as  non-active elliptical
populations.  More sophisticated  comparisons between isophotal, dusty
disk, and  radio jet axes may  yet reveal a  deeper connection between
these structures  than was previously thought  to exist.  Furthermore,
the  understanding of  dust in  radio-loud unification  models  may be
crucial  to  probing  the  divides  between FR  I  and  II-type  radio
galaxies,  and  whether  these   two  classes  reflect  stages  of  an
evolutionary   sequence,  or   are   intrinsically  distinct   objects
altogether.

\acknowledgments  We are  grateful  to Stefi  Baum,  David Floyd,  and
Andrew  McCullough for  helpful  discussions.  We  also thank  William
Keel,  our referee,  whose feedback  led  to the  improvement of  this
paper.  This work is based on observations made with the NASA/ESA {\it
Hubble  Space  Telescope}, obtained  at  the  Space Telescope  Science
Institute, which  is operated by  the Association of  Universities for
Research  in Astronomy,  Inc.,  under contract  NAS  5-26555 with  the
National Aeronautics and Space  Administration.  Support for this work
was provided  by NASA through grant HST-GO-10173.   This research made
extensive  use  of the  NASA  Astrophysics  Data System  Bibliographic
services, as well as the NASA/IPAC Extragalactic Database, operated by
the  Jet Propulsion  Laboratory, California  Institute  of Technology,
under contract with NASA.

\vspace{0.3in}


\LongTables


\begin{deluxetable}{lccccccccc}
\tabletypesize{\scriptsize}  
\tablecaption{Host Galaxy Isophotal Properties and Dust Distributions}
  \tablewidth{0pc}
  \tablehead{
    \colhead{Source} &
    \colhead{$\left( a_4/a \right)_{\mathrm{ave}}$} &
    \colhead{$\epsilon_{\mathrm{ave}}$} &
    \colhead{$z$} &
    \colhead{$n$} &
    \colhead{$r_{\mathrm{e}}/{\mathrm{kpc}}$} &
    \colhead{$m_{\mathrm{e}}$} &
    \colhead{Isophote Morphology} &
    \colhead{Dust Morphology} &
    \colhead{Reference} \\
    \colhead{(1)} & \colhead{(2)} & \colhead{(3)} &
    \colhead{(4)} & \colhead{(5)} & \colhead{(6)} &
    \colhead{(7)} & \colhead{(8)} & \colhead{(9)} &
    \colhead{(10)}
  }
  \startdata                                                                                     
3C~17 & 0.004313 & 0.19 & 0.21968 & 0.370 & 3.21 & 18.28 & Elliptical & --- & 1 \\
3C~20 & -0.06019 & 0.03 & 0.17400 & 0.350 & 2.33 & 17.89 & Boxy & Thin tendril & 1,2 \\
3C~28 & 0.034791 & 0.18 & 0.19520 & 0.821 & 1.05 & 17.76 & Disky & --- & 1 \\
3C~31 & -0.00041 & 0.11 & 0.01670 & 0.272 & 5.22 & 18.09 & Elliptical & Disk incl.~ $\sim 41^\circ$ & 1,2 \\
3C~33.1 & -0.03744 & 0.07 & 0.18090 & 0.443 & 3.08 & 18.65 & Boxy & --- & 1 \\
3C~35 & 0.007849 & 0.26 & 0.06700 & 0.371 & 5.89 & 18.77 & Elliptical & --- & 1 \\
3C~52 & 0.007006 & 0.30 & 0.28540 & 0.240 & 8.34 & 19.06 & Elliptical & Thin tendril & 1,2 \\
3C~61.1 & -0.001803 & 0.12 & 0.18400 & 0.282 & 2.86 & 19.52 & Elliptical & --- & 1 \\
3C~66B & 2.519e-06 & 0.06 & 0.02150 & 0.248 & 9.16 & 19.31 & Round & Face-on Disk & 1,2,3 \\
3C~75N & -0.000145 & 0.18 & 0.02315 & 0.247 & 1.98 & 17.10 & Round & --- & 1 \\
3C~76.1 & -0.002024 & 0.13 & 0.03240 & 0.261 & 1.08 & 16.70 & Round & Face-on Disk & 1,3 \\
3C~79 & -0.047640 & 0.19 & 0.25595 & 0.348 & 4.27 & 18.38 & Boxy & --- & 1 \\
3C~83.1 & -0.023451 & 0.29 & 0.02550 & 0.214 & 12.01 & 18.93 & Boxy & Edge-on Disk & 1,2 \\
3C~84 & $\sim -0.01$ 	& 0.22 & 0.01756 & 0.356 & 6.52 & 18.73 & --- & Complex patches & 1,2 \\
3C~88 & -0.046897 & 0.29 & 0.03022 & 0.227 & 4.92 & 19.18 & Boxy & Very thin, faint tendril & 1,2 \\
3C~98 & -0.010021 & 0.15 & 0.03000 & 0.248 & 3.12 & 18.31 & Boxy & --- & 1 \\
3C~105 &  $\sim -0.03$    &  0.28 & 0.08900 & 0.315 & 1.19 & 17.16 & --- & --- & 1 \\
3C~111 & -0.00032 & 0.09 & 0.04850 & 0.173 & 3.44 & 18.87 & Elliptical & --- & 1 \\
3C~123 & 0.008141 & 0.06 & 0.21770 & 0.166 & 28.65 & 22.28 & Elliptical & --- & 1 \\
3C~129 & -0.02592 & 0.14 & 0.02080 & 0.637 & 1.75 & 17.39 & Boxy & --- & 1 \\
3C~129.1 & 0.016238 & 0.06 & 0.02220 & 0.900 & 1.76 & 17.32 & Disky & --- & 1 \\
3C~132 & -0.002463 & 0.17 & 0.21400 & 0.362 & 5.36 & 18.66 & Elliptical & --- & 1 \\
3C~133 & 0.000983 & 0.06 & 0.27750 & 0.319 & 4.74 & 19.44 & Round & --- & 1 \\
3C~135 & -0.012420 & 0.19 & 0.12530 & 0.223 & 1.19 & 17.12 & Boxy & --- & 1 \\
3C~153 & -0.052845 & 0.10 & 0.27700 & 0.259 & 1.28 & 16.79 & Boxy & --- & 1 \\
3C~165 & 0.005589 & 0.13 & 0.29570 & 0.241 & 9.23 & 20.21 & Elliptical & --- & 1 \\
3C~166 & 0.075364 & 0.08 & 0.24500 & 0.225 & 14.18 & 21.49 & Disky & --- & 1 \\
3C~171 & -0.008469 & 0.06 & 0.23840 & 0.305 & 4.21 & 19.11 & Elliptical & Thin tendril & 1,2 \\
3C~173.1 & -0.049518 & 0.22 & 0.29210 & 0.162 & 9.01 & 19.46 & Boxy & Offset thin tendril & 1,2 \\
3C~180 &  -0.006289 & 0.38 & 0.22000 & 0.242 & 7.44 & 20.00 & Elliptical & Faint, thin patch & 1,2 \\
3C~184.1 & 0.0614536 & 0.25 & 0.11820 & 0.904 & 0.59 & 15.79 & Disky & --- & 1 \\
3C~192 & 0.008136 & 0.02 & 0.05980 & 0.253 & 2.61 & 17.84 & Elliptical & --- & 1 \\
3C~196.1 & -0.003252 & 0.28 & 0.19800 & 0.337 & 2.41 & 19.03 & Elliptical & Offset thin patch & 1,2 \\
3C~197.1 & -0.001396 & 0.04 & 0.13010 & 0.227 & 4.73 & 19.42 & Elliptical & --- & 1 \\
3C~198 & -0.070424 & 0.12 & 0.08150 & 0.292 & 3.50 & 19.31 & Boxy & --- & 1 \\
3C~213.1 & 0.004274 & 0.33 & 0.19400 & 0.281 & 4.59 & 18.95 & Elliptical & --- & 1 \\
3C~219 & 0.000479 & 0.20 & 0.17440 & 0.399 & 6.27 & 18.92 & Elliptical & --- & 1 \\
3C~223 & -0.006622 & 0.13 & 0.13680 & 0.286 & 4.06 & 18.71 & Elliptical & --- & 1 \\
3C~223.1 & 0.021664 & 0.32 & 0.10700 & 0.261 & 1.50 & 16.99 & Disky & Dust Lane(s) & 1 \\
3C~227 & 0.061402 & 0.12 & 0.08610 & 0.235 & 2.93 & 18.47 & Disky & --- & 1 \\
3C~234 & -0.027986 & 0.07 & 0.18500 & 0.433 & 2.78 & 17.88 & Boxy & --- & 1 \\
3C~236 & -0.069899 & 0.40 & 0.10050 & 0.291 & 10.06 & 19.57 & Boxy & Lane, Edge-on Disk & 1,2,4 \\
3C~264 & 0.000581 & 0.01 & 0.02172 & 0.671 & 1.52 & 16.13 & Round & Face-on Disk & 1,2 \\
3C~270 & -0.17046 & 0.27 & 0.00747 & --- & --- & --- & Boxy & Edge-on Disk & 2 \\
3C~277.3 & -0.023183 & 0.06 & 0.08570 & 0.255 & 5.54 & 19.11 & Boxy & --- & 1 \\
3C~284 & -0.021973 & 0.09 & 0.23900 & 0.213 & 6.13 & 19.44 & Boxy & Thin filament & 1,2 \\
3C~285 & -0.008023 & 0.25 & 0.07940 & 0.239 & 3.00 & 18.63 & Elliptical & --- & 1 \\
3C~287.1 & 0.066047 & 0.11 & 0.21590 & 0.268 & 1.00 & 16.24 & Disky & --- & 1 \\
3C~288 & -0.021296 & 0.03 & 0.24600 & 0.749 & 4.79 & 18.94 & Boxy & --- & 1 \\
3C~293 & ---  & --- & 0.04503 & --- & --- & --- & Highly irregular & Complex tendrils, lanes & 1,2,5 \\
3C~296 & -0.057085 & 0.28 & 0.02370 & 0.277 & 6.60 & 17.97 & Boxy & Edge-on Disk & 1 \\
3C~300 & -0.033492 & 0.25 & 0.27000 & 0.299 & 4.92 & 19.31 & Boxy & --- & 1 \\
3C~303 & -0.019399 & 0.09 & 0.14100 & 0.259 & 5.30 & 18.93 & Boxy & --- & 1 \\
3C~305 & $\sim -0.003$ & 0.35 & 0.04164 & 0.224 & 2.58 & 17.39 & Elliptical & Patches, tendrils & 1,2 \\
3C~310 & 0.0329299 & 0.25 & 0.05350 & 0.285 & 2.56 & 18.09 & Disky & --- & 1 \\
3C~314.1 & 0.000278 & 0.37 & 0.11970 & 0.408 & 4.44 & 18.99 & Elliptical & --- & 1 \\
3C~315 & 0.091022 & 0.33 & 0.10830 & 0.252 & 1.39 & 16.83 & Disky & --- & 1 \\
3C~317 & $\sim -0.01$ & 0.30 & 0.03446 & 0.434 & 1.63 & 17.83 & Elliptical & Thin tendrils & 1 \\
3C~319 & 0.027407 & 0.13 & 0.19200 & 0.353 & 0.78 & 16.40 & Disky & --- & 1 \\
3C~326 & 0.154769 & 0.42 & 0.08900 & 0.269 & 1.38 & 17.13 & Disky & Lane & 1,2 \\
3C~332 & -0.013333 & 0.06 & 0.15150 & 0.254 & 0.88 & 15.94 & Boxy & --- & 1 \\
3C~338 & $\sim -0.001$	& 0.28 & 0.03035 & 0.474 & 12.86 & 19.03 & Round & Tendrils, clumps & 1,2 \\
3C~346 & -0.004300 & 0.24 & 0.16100 & 0.270 & 0.46 & 14.90 & Elliptical & --- & 1 \\
3C~348 & -0.002189 & 0.17 & 0.15400 & 0.925 & 9.25 & 19.58 & Elliptical & --- & 1 \\
3C~349 & -0.061500 & 0.40 & 0.20500 & 0.249 & 1.48 & 17.27 & Boxy & --- & 1 \\
3C~353 & 0.019501 & 0.02 & 0.03043 & 0.251 & 1.85 & 17.86 & Disky & --- & 1 \\
3C~357 & 0.047445 & 0.29 & 0.16700 & 0.238 & 5.29 & 18.94 & Disky & Faint, thin tendril(s) & 1,2 \\
3C~379.1 & 0.011905 & 0.07 & 0.25600 & 0.225 & 2.77 & 18.14 & Disky & --- & 1 \\
3C~381 & -0.005648 & 0.17 & 0.16050 & 0.264 & 1.56 & 17.10 & Elliptical & --- & 1 \\
3C~386 & 0.014887 & 0.12 & 0.01700 & 0.226 & 3.38 & 18.32 & Disky & --- & 1 \\
3C~388 & 0.016990 & 0.13 & 0.09100 & 1.227 & 2.21 & 17.64 & Disky & --- & 1 \\
3C~401 & -0.019726 & 0.13 & 0.20104 & 0.310 & 1.17 & 17.89 & Boxy & --- & 1 \\
3C~402 & 0.005824 & 0.18 & 0.02390 & 0.190 & 5.00 & 18.85 & Elliptical & --- & 1 \\
3C~403 & 0.034994 & 0.27 & 0.05900 & 0.405 & 4.72 & 18.01 & Diksy & Lane & 1,2 \\
3C~405 & --- & 0.25 & 0.05608 & 0.525 & 14.85 & 20.01 & Irregular, obscured & Complex, offset patch & 1,2 \\
3C~424 & -0.014721 & 0.08 & 0.12699 & 0.254 & 3.67 & 19.38 & Boxy & --- & 1 \\
3C~430 & 0.0475771 & 0.32 & 0.05560 & --- & --- & --- & Diksy & Lane & 2 \\
3C~433 & -0.003694 & 0.37 & 0.10160 & 0.494 & 2.51 & 17.90 & Elliptical & Patchy clumps & 1,2 \\
3C~436 & -0.035997 & 0.22 & 0.21450 & 0.255 & 6.61 & 19.12 & Boxy & Edge-on Disk & 1,2 \\
3C~438 & 0.011698 & 0.05 & 0.29000 & 0.298 & 9.96 & 19.86 & Disky & --- & 1 \\
3C~449 & -0.004330 & 0.15 & 0.01710 & 0.164 & 5.08 & 19.01 & Elliptical & Warped disk incl. $\sim 40^\circ$ & 1,2,6 \\
3C~452 & -0.016480 & 0.27 & 0.08110 & 0.276 & 3.83 & 18.15 & Boxy & Edge-on Disk & 1,2 \\
3C~459 & -0.000975 & 0.12 & 0.21990 & 0.430 & 0.81 & 15.45 & Elliptical & --- & 1 \\
3C~465 & -0.003960 & 0.16 & 0.03030 & 0.311 & 3.92 & 17.67 & Elliptical & Face-on Disk & 1,2 \\
NGC 6251 & -0.183990 & 0.17 & 0.02471 & --- & --- & --- & Boxy & Edge-on Disk & 7 \\    
  \enddata
  \tablecomments{
    (1) Source name;
    (2) 4th order cosine coefficient of {\sc ellipse} fit normalized
to the isophote semi-major
axis $a$. Negative values of $a_4/a$ are representative of boxy isophotes, whereas
positive values indicate a disky profile; 
    (3) Isophotal ellipticity averaged over the same region as in (2);
    (4) Host galaxy redshift;
    (5) S\'{e}rsic index of model;
    (6) Effective radius of model in kpc;
    (7) Effective host surface brightness in mag arcsec$^{-2}$, normalized to $H$-band Vega magnitudes;
    (8) Qualitative isophotal morphology;
    (9) Qualitative description of dust content and distribution, where applicable. No entry indicates 
that the host is optically dust-poor. (``Face-on'' and ``Edge-on'' is a strictly qualitative assertion);
    (10) References
  }
  \tablerefs{
(1)~\citet{donzelli06}; 
(2)~\citet{dekoff00}; 
(3)~\citet{sparks00}; 
(4)~\citet{odea01}; 
(5)~\citet{floyd06};
(6)~\citet{tremblay06};
(7)~\citet{ferrarese99}}
\label{tab:tab1}
\end{deluxetable}




\end{document}